\documentclass[11pt,a4paper]{article}
\usepackage{jheppub}

\usepackage{amsmath}
\usepackage{graphicx}
\usepackage{url}
\usepackage{epsfig}
\usepackage{feynmf}

\newcommand{\sq}{{\tilde{q}}}
\newcommand{\sqb}{{\bar{\tilde{q}}}}
\newcommand{\gl}{{\tilde{g}}}
\renewcommand{\d}{\mathrm{d}}

\title{Towards NNLL resummation: hard matching coefficients for squark and gluino hadroproduction}
\author[a]{Wim Beenakker,}
\author[a]{Tim Janssen,}
\author[a]{Susanne Lepoeter,}
\author[b]{Michael Kr\"amer,}
\author[c]{Anna Kulesza,}
\author[d]{Eric Laenen,}
\author[a]{Irene Niessen,}
\author[e,1]{Silja Thewes%
\note{former address},}
\author[a]{Tom Van Daal}

\affiliation[a]{Theoretical High Energy Physics, IMAPP, Faculty of Science, Mailbox 79, Radboud University Nijmegen, P.O. Box 9010, NL-6500 GL Nijmegen, The Netherlands}
\affiliation[b]{Institute for Theoretical Particle Physics and Cosmology, RWTH Aachen University D-52056 Aachen, Germany}
\affiliation[c]{Institute for Theoretical Physics, WWU M\"unster, D-48149 M\"unster, Germany}
\affiliation[d]{ITFA, University of Amsterdam, Science Park 904, 1018 XE, Amsterdam,\\
ITF, Utrecht University, Leuvenlaan 4, 3584 CE Utrecht,\\
Nikhef Theory Group, Science Park 105, 1098 XG Amsterdam, The Netherlands}
\affiliation[e]{DESY, Theory Group, Notkestrasse 85, D-22603 Hamburg, Germany}

\emailAdd{W.Beenakker@science.ru.nl}
\emailAdd{silja.christine.thewes@desy.de}
\emailAdd{T.H.W.Janssen@student.ru.nl}
\emailAdd{S.Lepoeter@student.ru.nl}
\emailAdd{mkraemer@physik.rwth-aachen.de}
\emailAdd{anna.kulesza@uni-muenster.de}
\emailAdd{t45@nikhef.nl}
\emailAdd{i.niessen@science.ru.nl}
\emailAdd{TomvanDaal@student.ru.nl}

\abstract{We present the hard matching coefficients for squark and gluino hadroproduction. The hard matching coefficients follow from the next-to-leading order cross section near threshold and are an important ingredient for performing threshold resummation at next-to-next-to-leading logarithmic accuracy. We discuss the calculation, list the analytical results and study the numerical impact of these corrections. We find that the impact of the hard matching coefficients can be considerable, with the largest effect observed for final states involving gluinos.}


\preprint{DESY 13-072, MS-TP-13-10, Nikhef-2012-016,\\ 
\hspace*{13cm} TTK-13-10}

\begin{document}
\maketitle

\begin{fmffile}{feyngraphs}
\fmfset{dot_size}{1thick}
\fmfset{curly_len}{1.5mm}
\fmfset{arrow_len}{3mm}
\fmfpen{thin}
\setlength{\unitlength}{1mm}

\section{Introduction}
\label{s:intro}

Weak-scale supersymmetry (SUSY) \cite{Golfand:1971iw,Wess:1974tw} predicts new particles with masses in the TeV range, which could be detected at the Large Hadron Collider (LHC). In particular the coloured squarks ($\tilde q$) and gluinos ($\tilde g$) would be produced abundantly in hadronic collisions. Searches for squarks and gluinos at the LHC have placed lower limits on squark and gluino masses around 1~TeV \cite{Aad:2012rz,Aad:2012ona,Chatrchyan:2012gq,Chatrchyan:2012jx}. Once the LHC reaches its design energy near $\sqrt{S}=14$~TeV, SUSY particles with masses up to 3~TeV can be probed \cite{Aad:2009wy,Bayatian:2006zz}.

In the context of the Minimal Supersymmetric Standard Model (MSSM) \cite{Nilles:1983ge,Haber:1984rc} with $R$-parity conservation, squarks and gluinos are produced in pairs in collisions of two hadrons $h_1$ and $h_2$:
\begin{equation}
  h_1 h_2 \;\to\; \tilde{q}\tilde{q}\,,
  \tilde{q}\sqb\,, \tilde{q}\tilde{g}\,, \tilde{g}\tilde{g} + X\,.
\label{eq:processes}
\end{equation}
In Eq.~(\ref{eq:processes}) and throughout the rest of this paper we
suppress the chiralities of the squarks $\tilde{q} =(\tilde{q}_{L},
\tilde{q}_{R})$ and do not explicitly state the charge-conjugated
processes.  We include squarks $\sq$ of any flavour except for top
squarks.  The production of top squarks~\cite{Beenakker:1997ut} has to
be considered separately since the strong Yukawa coupling between top
quarks, top squarks and Higgs fields gives rise to potentially large
mixing effects and mass splitting~\cite{Ellis:1983ed}.
Accurate theoretical predictions for inclusive squark and gluino cross sections are needed to set exclusion limits and to help determining SUSY particle masses and properties in case of discovery \cite{Baer:2007ya,Dreiner:2010gv}. The precision of the predictions can be improved considerably by including higher-order SUSY-QCD corrections, which have been known for quite some time at next-to-leading order (NLO) in SUSY-QCD \cite{Beenakker:1994an,Beenakker:1995fp,Beenakker:1996ch}. A significant part of these corrections can be attributed to the threshold region, where the partonic centre-of-mass (CM) energy is close to the kinematic production threshold. The NLO corrections are then dominated by soft-gluon emission off the coloured particles in the initial and final state and by the Coulomb corrections due to the exchange of gluons between the slowly moving massive particles in the final state. The dominant contributions due to soft-gluon emission have the general form
\begin{equation}
\alpha_{\rm s}^n \log^m\!\beta^2\ \ , \ \ m\leq 2n 
\qquad {\rm \ with\ } \qquad 
\beta^2 \,\equiv\, 1-\rho\equiv1 \,-\, \frac{4m_{av}^2}{s}\,,
\label{eq:beta}
\end{equation}
where $\alpha_{\rm s}$ is the strong coupling, $s$ is the partonic CM energy squared and $m_{av}$ is the average mass of the final-state particles. These soft-gluon corrections can be taken into account to all orders in perturbation theory by means of threshold resummation techniques \cite{Sterman:1986aj,Catani:1989ne,Bonciani:1998vc,Contopanagos:1996nh,
  Kidonakis:1998bk,Kidonakis:1998nf}. The all-order summation of such logarithmic terms is a consequence of the near-threshold factorization of the cross section. We perform the resummation after taking a Mellin transform (indicated by a tilde) of the hadronic cross section $\sigma_{h_1h_2\to kl}$:
\begin{align}
  \label{eq:10}
  \tilde\sigma_{h_1 h_2 \to kl}\bigl(N, \{m^2\}\bigr) 
  &\equiv \int_0^1 d\rho\;\rho^{N-1}\;
           \sigma_{h_1 h_2\to kl}\bigl(\rho,\{ m^2\}\bigr)\,,
           \end{align}
with $k$ and $l$ the final-state particles and $\{m^2\}$ the squared masses involved in the process.
The logarithmically enhanced terms are then of the form $\alpha_{\rm
  s}^n \log^m N$, $m\leq 2n$, with the threshold limit
$\beta\rightarrow 0$ corresponding to $N\rightarrow \infty$. 

Threshold resummation has been performed for all MSSM squark and gluino production processes at next-to-leading-logarithmic (NLL) accuracy \cite{Kulesza:2008jb,Kulesza:2009kq,
Beenakker:2009ha,Beenakker:2010nq,Beenakker:2011fu}. For squark-antisquark and gluino-pair production, in addition to soft-gluon resummation, the Coulomb corrections have been resummed both by using a Sommerfeld factor \cite{Kulesza:2009kq} and by employing the framework of soft-collinear effective field theories \cite{Beneke:2010da,Falgari:2012hx,Falgari:2012sq}. Furthermore, the
dominant next-to-next-to-leading order (NNLO) corrections, including those coming from the resummed cross section at
next-to-next-to-leading-logarithmic (NNLL) level, have been calculated for squark-antisquark and gluino pair-production \cite{Langenfeld:2009eg,langenfeldgluino}. Very recently, approximate NNLO predictions for stop-antistop  production have been obtained using the soft-collinear effective field theory formalism~\cite{Broggio:2013uba}. For squark-antisquark production, soft-gluon emissions have been resummed to NNLL accuracy \cite{Beenakker:2011sf}, resulting in a notable stabilization of the theoretical predictions. Recently, the NNLL resummation has been also performed for gluino-pair production~\cite{Pfoh:2013iia}\footnote{However, the result in \cite{Pfoh:2013iia} relies on results of~\cite{Kauth:2011vg} which we comment on in section~\ref{s:Ccoeff}.}. One would expect similar results for the other SUSY-QCD production channels. In the following, we discuss the calculation of a particular class of ingredients necessary for performing NNLL resummation, the so-called hard matching coefficients.

The squark and gluino production processes listed in Eq.~(\ref{eq:processes}) are scattering processes with a non-trivial colour structure. Since the soft radiation is coherently sensitive to the colour structure of the underlying hard scattering, colour correlations have to be taken into account when considering resummation. In an appropriately chosen colour basis, the NNLL resummed partonic cross section $\sigma_{ij\to kl}^{\rm (res)}$ takes the form of a sum over irreducible representations $I$~\cite{Bonciani:1998vc, Beneke:2010da}:
\begin{align}
  \label{eq:12}
  \tilde{\sigma}^{\rm (res)} _{ij\rightarrow kl}\bigl(N,&\{m^2\},\mu^2\bigr) 
  =\sum_{I}\tilde\sigma^{(0)}_{ij\rightarrow kl,I} \bigl(N,\{m^2\},\mu^2\bigr)\\
  &\times\left(1+\frac{\alpha_{\rm s}}{\pi}\;{\cal C}^{\rm Coul,(1)}_{ij\rightarrow kl,I}(N,\{m^2\},\mu^2)\right)\left(1+\frac{\alpha_{\rm s}}{\pi}\;{\cal C}^{\rm (1)}_{ij\rightarrow kl,I}(\{m^2\},\mu^2)\right)\nonumber\\
  & \times\Delta_i (N+1,Q^2,\mu^2)\,\Delta_j (N+1,Q^2,\mu^2)\,
     \Delta^{\rm (s)}_{ij\rightarrow kl,I}\bigl(Q/(N\mu),\mu^2\bigr)\,.\nonumber
\end{align}
Here $\tilde{\sigma}^{(0)}_{ij \rightarrow kl,I}$ is the colour-decomposed leading order (LO) partonic cross section in Mellin-moment space, $\mu$ is the common factorization and renormalization scale, and we introduced the hard scale $Q^2 = 4m_{av}^2$. The last line of Eq.~\eqref{eq:12} captures all dependence on the large logarithm. The functions $\Delta_{i}$ and $\Delta_{j}$ sum the effects of the (soft-)collinear radiation from the incoming partons, while the function $\Delta^{\rm (s)}_{ij\rightarrow kl,I}$ describes the wide-angle soft radiation. These functions are known at NNLL, see e.g. \cite{Beenakker:2011sf} and references therein. The matching coefficients in the second line contain the Mellin moments of the higher-order contributions without the $\log(N)$ terms. This non-logarithmic part of the higher-order cross section near threshold factorizes into a part that contains the leading Coulomb correction ${\cal C}^{\rm Coul,(1)}_{ij\rightarrow kl,I}$ and a part that contains the NLO hard matching coefficients ${\cal C}^{\rm (1)}_{ij\rightarrow kl,I}$~\cite{Bonciani:1998vc,Beneke:2010da}. These hard matching coefficients follow from the threshold limit of the full NLO calculation. They are a key element in the NNLL calculation and will be the subject of this paper.
 
We will present the colour-decomposed NLO hard matching coefficients for the squark and gluino production processes at the LHC and discuss the impact of the corrections. We will start by constructing an appropriate colour basis in SU($N_c$) in section~\ref{s:colour} and discuss why some colour-decomposed cross sections vanish at threshold in section~\ref{s:zero}. In section~\ref{s:Ccoeff} we discuss the calculation of the matching coefficients. The numerical results are presented in section~\ref{s:numres} where we show predictions for the LHC with centre-of-mass energy of $\sqrt{S}=8$~TeV. We conclude in section~\ref{s:conclusion}.

\section{Construction of the colour bases in SU($\bf N_c$)}
\label{s:colour}

When performing resummation for coloured particles, one has to take into account the colour correlations introduced by gluon radiation. In particular, wide-angle soft radiation
is coherently sensitive to the colour structure of the hard process
from which it is emitted \cite{Bonciani:1998vc,Contopanagos:1996nh,Kidonakis:1998bk,Kidonakis:1998nf,Botts:1989kf,Kidonakis:1997gm}. At threshold, the resulting colour matrices become diagonal to all orders by performing the calculation in an $s$-channel colour basis \cite{Beneke:2009rj,Kulesza:2008jb,Kulesza:2009kq}. This basis traces the colour flow through the $s$-channel and is obtained by performing an $s$-channel colour decomposition of the reducible two-particle product representations into irreducible ones. Methods to obtain such a basis have been presented in \cite{Beenakker:2009ha} and \cite{Beneke:2009rj} for processes that contain particles of the same representations in the initial and final state and recently for any number of partons in SU($N_c$) in Ref.~\cite{Keppeler:2012ih}. Here we will present a method that is also valid for processes that contain particles of different representations in the initial and final state. All results are shown for a general SU($N_c$) theory with $N_c$ the number of colours.

We denote the colour charge operator of a representation by $T_j$. It is given by gluon emission off the corresponding particle $j$, so 
\begin{align}
(T_{q,\sq})^c_{ab}=T^c_{ab},\quad (T_{\bar q,\sqb})^c_{ab} =-T^c_{ba}=(T^c_{ab})^*\quad\mbox{and}\quad (T_{g,\gl})^c_{ab}=F^{c}_{ab}=-if_{abc}\,,
\end{align}
with $T^c_{ab}$ and $F^c_{ab}$ the generators of the fundamental and the adjoint representation respectively. In addition to the completely antisymmetric structures $F^c_{ab}$, we will also need the traceless symmetric octet structures $D^c_{ab}=d_{abc}$ and the singlet colour structures $\delta_{ab}$. The dimension of the colour labels is determined by the particle they refer to. In SU($N_c$), the quarks and squarks are in the $\bf N_c$-dimensional fundamental representation, while the gluons and gluinos are in the adjoint representation with dimension $\bf N_c^2-1$. The colour decompositions for the squark and gluino production processes are given by:
\begin{align}
q\bar q\to\sq\sqb: &\quad\mathbf{1}\oplus\mathbf{(N_c^2-1)}\,,\label{eq:qqbtossbcoldec}\\[2mm]
gg\to\sq\sqb:&\quad\mathbf{1}\oplus\mathbf{(N_c^2-1)_A}\oplus\mathbf{(N_c^2-1)_S} \,,\\[2mm]
q\bar q\to\gl\gl:&\quad\mathbf{1}\oplus\mathbf{(N_c^2-1)_A}\oplus\mathbf{(N_c^2-1)_S}\,,\\[2mm]
gg\to\gl\gl:&\quad\mathbf{1}\oplus\mathbf{(N_c^2-1)_A}\oplus(\mathbf{N_c^2-1)_S}\oplus\mathbf{(N_c^2-1)(N_c^2-4)/4}\oplus\overline{\mathbf{(N_c^2-1)(N_c^2-4)/4}}\nonumber\\
&\qquad\oplus\mathbf{N_c^2(N_c+3)(N_c-1)/4}\oplus\mathbf{N_c^2(N_c-3)(N_c+1)/4}\,,\\[2mm]
qq\to\sq\sq:&\quad\mathbf{N_c(N_c-1)/2}\oplus\mathbf{N_c(N_c+1)/2}\,, \label{eq:ggtoggcoldec}\\[2mm]
qg\to\sq\gl:& \quad\mathbf{N_c}\oplus\overline{\mathbf{N_c(N_c+1)(N_c-2)/2}} \oplus\mathbf{N_c(N_c-1)(N_c+2)/2}\,.\label{eq:qgtosgcoldec}
\end{align}
In SU$(3)$, the $N_c(N_c-1)/2$-dimensional representation for the $qq\to\sq\sq$ process coincides with the antifundamental representation $\overline{\bf 3}$, while the last representation of the $gg\to\gl\gl$ process vanishes.

An $s$-channel colour basis can be constructed from any complete basis by requiring orthogonality, normalization and that it is an eigenvector under the quadratic Casimir operator. The orthogonality and normalization follow from an inner product of two colour tensors $c_I$ and $c_J$ that describe the colour content of the $2\to2$ process under consideration:
\begin{align}
c_I\cdot c_J\equiv\parbox{34mm}{\begin{fmfchar*}(34,18)
\fmfleft{dl,ul}
\fmfright{dr,ur}
\fmf{plain,label=\footnotesize$a_2$,label.dist=8,label.side=left}{dl,v}
\fmf{plain,label=\footnotesize$a_1$,label.dist=8,label.side=left}{v,ul}
\fmf{plain}{ul,ur,v1,dr,dl}
\fmfv{decor.shape=circle,decor.filled=empty,decor.size=6mm,label=$c_I$,label.dist=0}{v}
\fmf{plain,left,label=\footnotesize$a_3$,label.dist=8}{v,v1}
\fmf{plain,right,label={\footnotesize$a_4$},label.dist=8}{v,v1}
\fmfv{decor.shape=circle,decor.filled=empty,decor.size=6mm,label=$c_J$,label.angle=45,label.dist=0}{v1}
\fmfforce{(0.58w,0.53h)}{star}
\fmfv{label=$\!^*$}{star}
\end{fmfchar*}}=c_I(a_1,a_2,a_3,a_4)\,c_J^*(a_1,a_2,a_3,a_4)=\mbox{dim}(c_I)\,\delta_{IJ}\,,\label{eq:coltensornorm}
\end{align}
where $a_1$ and $a_2$ are the colour labels of the initial-state particles and $a_3$ and $a_4$ the colour labels of the final-state particles.
The last equality in Eq.~\eqref{eq:coltensornorm} fixes the normalization as well as the orthogonality. In addition, base tensors have to be eigenvectors of the quadratic Casimir operator $(T_i+T_j)^2$, with $T_i$ and $T_j$ the colour charge operators of the two initial-state particles:
\begin{align*}
\Big((T_i)^c_{a_1d_1}\delta_{a_2d_2}+(T_j)^c_{a_2d_2}\delta_{a_1d_1}\Big)&\Big((T_i)^c_{d_1b_1}\delta_{d_2b_2}+(T_j)^c_{d_2b_2}\delta_{d_1b_1}\Big)\delta_{a_3b_3}\delta_{a_4b_4}c_I(b_1,b_2,b_3,b_4)\\
&=C_2(R_I)c_I(a_1,a_2,a_3,a_4)\,,
\end{align*}
or in a shorter notation:
\begin{align}
(T_i+T_j)^2\otimes c_I=C_2(R_I)c_I\,.\label{eq:casimireq}
\end{align}
Here $C_2(R_I)$ is the quadratic Casimir invariant of the representation $R_I$ that corresponds to the base tensor $c_I$. Eq.~\eqref{eq:casimireq} provides both the value of the quadratic Casimir invariant and an additional constraint for the base tensor.

The combined requirements of orthogonality, normalization and invariance under the quadratic Casimir operator fix the base tensors up to a phase. The resulting base tensors and their corresponding dimension and quadratic Casimir invariant for SUSY-QCD processes are listed in Appendix~\ref{app:colourstructures}.

\section{Suppressed cross sections}\label{s:zero}

Near threshold, the $s$-wave contribution to the final state dominates. Higher values of the final-state orbital angular momentum quantum number $L_{\rm fin}$ are suppressed by higher powers of $\beta$. Thus a cross section can be regarded as being ``suppressed" near threshold if the $L_{\rm fin}=0$ mode is not accessible due to symmetry properties. In this section, we will first list these general symmetry considerations and then apply them to the different processes.

If the initial or final state consists of two identical particles, the symmetry  properties are determined by the eigenvalue $P$ of the permutation operator that interchanges the two particles:
\begin{equation}
P={\cal S}_c\,(-1)^{L+S-S_1-S_2}=\bigg\{\begin{array}{ll}+1\quad&\rm{for\;identical\;bosons,}\\-1&\rm{for\;identical\;fermions.}\end{array}\label{eq:symmetry}
\end{equation}
Here $L$ is the total orbital angular momentum of the particle pair in the CM frame, $S$~the total spin, $S_1$ and $S_2$ the spins of the individual particles and ${\cal S}_c$ the colour symmetry factor, which is $+1$ for a symmetric colour state and $-1$ for an antisymmetric colour state. Because we have $L_{\rm fin}=0$ for a non-suppressed cross section, the conserved angular momentum quantum number equals the final-state spin, i.e. $J=S_{\rm fin}$. In addition, we will use the conserved total angular momentum quantum number $M_J\in \{-J,\ldots,J\}$ for quantization along the propagation axis of the incoming massless partons ($z$-axis). In that case $M_J$ is given by the difference of the helicities of the initial-state partons.  

\paragraph{\boldmath$q_iq_i\to\sq_i\sq_i$: only the symmetric colour structure contributes.} We will argue that if the produced squarks in the $qq\to\sq\sq$ process have the same flavour, the contribution from the antisymmetric colour structure is suppressed near threshold. In that case, we have a system with identical fermions in the initial state with $P=-1$ and identical bosons in the final state with $P=+1$.  Also, since squarks are scalars, the total spin of the final state is $S_{\rm fin}=0$. We need $L_{\rm fin}=0$ for a non-suppressed threshold cross section, so the conserved angular momentum quantum number is $J=0$, which automatically means $L_{\rm in}=S_{\rm in}$. Inserting this into Eq.~\eqref{eq:symmetry} for the initial-state quarks, we see that the exponent is always odd. Eq.~\eqref{eq:symmetry} shows that we can only have an antisymmetric state if ${\cal S}_c=+1$, i.e. the colour structure is symmetric. Indeed, as we set out to argue, for equal flavours the antisymmetric colour structure is suppressed near threshold. 

This latter statement can be derived even more directly from the final state, which should be symmetric since squarks are spin-0 particles. Given that $S_{\rm fin}=S_1=S_2=0$ and that we want $L_{\rm fin}=0$, one readily obtains the requirement that ${\cal S}_c=+1$ from Eq.~\eqref{eq:symmetry}. 

\paragraph{\boldmath$gg\to\sq\sqb$: only the symmetric colour structures contribute.} In a similar way as in the first argument for $q_iq_i\to\sq_i\sq_i$, we can show that for the $gg\to\sq\sqb$ process only the symmetric colour structures contribute near threshold. In this case the gluons need to be in a symmetric state, since they are spin-1 particles. As with the previous process, we have $J=0$ based on the final state. Using Eq.~\eqref{eq:symmetry} we can see that the exponent is always even, so also in this case only the symmetric colour states can yield a non-suppressed contribution near threshold.

\paragraph{\boldmath$gg\to\gl\gl$: only the symmetric colour structures contribute.}
First, note that for the production of a pair of gluinos the final state should be antisymmetric, since gluinos are spin-$\tfrac{1}{2}$ particles. Given the fact that we need $L_{\rm fin}=0$ (see above) and that $J=S_{\rm fin}=0$ or $1$, Eq.~\eqref{eq:symmetry} leads to the requirement that $(-1)^{J-1}{\cal S}_c=-1$. So, if the gluinos are produced in a $J=0$ ($J=1$) state the colour structure should have ${\cal S}_c=+1 (-1)$, i.e. be symmetric (antisymmetric). Which of these situations is realized will depend on the initial state. For the $gg\to\gl\gl$ process, the gluons have helicities $\pm 1$ and the total angular momentum quantum number for quantization along the $z$-axis is accordingly given by $M_J=\pm 2,0$. The case $M_J=0$ corresponds to gluons with equal helicities and the case $M_J=\pm 2$ implies $J\ge 2$, which is ruled out by the final-state requirement that $J=0$ or 1. This $J$-ambiguity we must now settle.\\[1mm]
At this point we would have liked to invoke the Landau-Yang theorem~\cite{LandauYang} to rule out the $J=1$ option. After all, in the threshold limit the gluino pair is produced at rest and can therefore effectively be regarded as a spin-$J$ particle that (in the time-reversed sense) is decaying into two massless spin-1 gluons. However, if we carefully follow the argumentation of the Landau-Yang theorem, which is formulated for photons rather than gluons, no additional constraint is obtained. The colour quantum numbers of the gluons provide in fact a loophole for evading the implications of the Landau-Yang theorem. \\[1mm]
Nevertheless in our results we observe that only the symmetric colour states yield a non-suppressed threshold contribution to the LO and NLO cross sections, implying that only the $J=0$ state survives. The reason for this turns out to reside in the LO matrix element, which also features prominently in the calculation of the NLO virtual corrections. This LO matrix element can be cast into a multiplicative form consisting of two factors, one containing the colour structure and one containing all Lorentz and spin structure. In view of the gluino-pair symmetry arguments presented above, this means that only symmetric or antisymmetric colour structures can survive in the threshold limit and not both types of structures simultaneously.

\paragraph{\boldmath$q\bar q\to\gl\gl$: only the antisymmetric colour structure contributes.} For the $q\bar q\to\gl\gl$ process we can make use of the fact that the helicities of the massless initial-state quark and antiquark are opposite, as a result of chirality conservation in SUSY-QCD.  The case of equal helicities is suppressed by the negligible mass of the initial-state quarks. Therefore, $M_J=\pm 1$ for angular momentum quantization along the $z$-axis, which excludes the $J=0$ gluino-pair state. Based on the gluino-pair symmetry arguments presented earlier, this requires the gluinos to be in an ${\cal S}_c=-1$ state, i.e. an antisymmetric colour state, as is indeed observed.\\ 

These symmetry arguments remain also valid at higher orders, so the hard matching coefficients of suppressed cross sections only contribute to $1/N$-suppressed terms.\footnote{An exception might be the $gg\to\gl\gl$ case, which
                might receive a contribution from the antisymmetric
                colour structures at NNLO level from squaring the NLO
                matrix element. However, the hard matching coefficients
                at NLO are not affected by this.}
Since these higher-order $1/N$-suppressed effects are beyond the level of accuracy of resummation we want to perform, we put the value of the hard matching coefficients for the suppressed cross sections to zero. Note however, that we choose to keep the $1/N$-suppressed terms originating from the LO cross sections in Eq.~(\ref{eq:12}).

\section{Calculation of the hard matching coefficients}
\label{s:Ccoeff}

In this section we will discuss the calculation of the hard matching coefficients ${\cal C}^{\rm (1)}$ at one loop. The hard matching coefficients ${\cal C}^{\rm(1)}$ are determined by the terms in the $\beta$-expansion of the NLO cross section that are proportional to $\beta$,  $\beta \log(\beta)$ and  $\beta \log^2(\beta)$. These terms receive contributions from both the real and the virtual corrections. After taking a Mellin transform, terms that contain higher powers of $\beta$ are suppressed by powers of $1/\sqrt{N}$ and do not have to be considered.

To obtain the integrated real corrections at threshold, we observe that they are formally phase-space suppressed near threshold unless the integrand compensates this suppression. Therefore we can construct the real corrections at threshold from the singular behaviour of the matrix element squared, which can be obtained using dipole subtraction \cite{Catani:1996vz,Catani:2002hc}. This has been discussed in detail in Ref.~\cite{Beenakker:2011sf} in the context of squark-antisquark production. The same procedure can be used for the other SUSY-QCD processes and the result in $n=4-2\varepsilon$ dimensions is given by:
\begin{align}
\sigma^{\rm R,thr}_I&=16\pi\alpha_{\rm s}S_n\sigma^{\rm LO,thr}_I\Bigg\{C_2(R_I)\left(\frac{1}{2\varepsilon}-\log(8\beta^2)+3\right)\label{eq:sigmareal}\\
&\quad+\sum_{n=\{i,j\}}T_n^2\left[\frac{1}{2\varepsilon^2}-\frac{1}{\varepsilon}\left(\log(2)-\frac{\gamma_n}{2T_n^2}\right)+\log^2(8\beta^2)-4\log(8\beta^2)\right.\nonumber\\
&\left.\qquad+8-\frac{11\pi^2}{24}-\log\left(\frac{\mu_F^2}{m_{av}^2}\right)\left(\log(8\beta^2)-2+\frac{\gamma_n}{2T_n^2}-\log(2)\right)\right]\Bigg\}.\nonumber
\end{align}
Thus the real threshold cross section $\sigma^{\rm R,thr}_I$ corresponding to a representation $I$ is proportional to the colour-decomposed LO cross section at threshold $\sigma^{\rm LO,thr}_I$. The final-state contributions are weighted by the quadratic Casimir invariant $C_2(R_I)$ of the representation $R_I$, which is listed in Appendix~\ref{app:colourstructures}. The sum in the last two lines of Eq.~\eqref{eq:sigmareal} runs over the initial-state partons. The colour operators $T_n^2$ depend on the representation of the corresponding particle, while the values of the flavour coefficients $\gamma_n$ are determined by the partons in the initial state:
\begin{align*}
T_{q}^2=C_F=\frac{N_c^2-1}{2N_c}\,,\quad T_{g}^2=C_A=N_c\,,\quad\gamma_q=\frac{3}{2}C_F\,,\quad\gamma_g=\frac{11}{6}C_A-\frac{1}{3}n_l
\end{align*}
where $n_l=5$ is the number of light flavours. Finally, $\mu_F$ is the factorization scale and the factor $S_n$ is given by:
\[S_n=\frac{1}{16\pi^2}e^{-\varepsilon(\gamma_E-\log(4\pi))}\left(\frac{\mu_R^2}{m_{av}^2}\right)^{\varepsilon}\]
with $\gamma_E$ being Euler's constant and $\mu_R$ the renormalization scale. Eq.~\eqref{eq:sigmareal} is valid for all pair-production processes in SUSY-QCD that have a linear behaviour in $\beta$ at LO, including the unequal mass case of squark-gluino production.

To obtain the virtual corrections, we start from the full analytic calculation as presented in Ref.~\cite{Beenakker:1996ch}. As described in detail in Ref.~\cite{Beenakker:1996ch}, the QCD coupling $\alpha_{\rm s}$ and the parton distribution functions at NLO are defined in the $\overline{\rm MS}$ scheme with five active flavours, with a correction for the SUSY breaking in the $\overline{\rm MS}$ scheme. The masses of squarks and gluinos are renormalized in the on-shell scheme, and the top quark and the SUSY particles are decoupled from the running of $\alpha_{\rm s}$.

We first need to colour-decompose the virtual corrections and then expand them in $\beta$. These steps have been discussed for squark-antisquark production in Ref.~\cite{Beenakker:2011sf} and the same procedure can be followed for the other processes although for the processes involving gluinos one does need to pay attention to spurious $1/\beta$ singularities originating from Gram determinants occurring in the reduction of tensor integrals. After expanding to sufficient powers in $\beta$, these singularities cancel in the calculation. The hard matching coefficients are now obtained by adding the non-logarithmic terms of the Mellin transform of Eq.~\eqref{eq:sigmareal} to the Mellin transform of the ${\cal O}(\beta)$ virtual corrections to the cross section and dividing the result by the LO threshold cross section in Mellin-moment space. Note that the NLO hard matching coefficients will thus also contain non-logarithmic contributions from Mellin transforms of the ${\cal O}(\beta)$ logarithmic terms in Eq.~\eqref{eq:sigmareal}. The complete expressions for the hard matching coefficients of the SUSY-QCD processes can be found in Appendix~\ref{app:Ccoeff}. The virtual part of the corrections to the $gg\to\gl\gl$ process agree with the results presented in~\cite{Kauth:2011vg}, provided that one translates their $\overline{\rm DR}$ result to our $\overline{\rm MS}$ result and manually decouples the heavy particles from the running of $\alpha_{\rm s}$. On top of that, the top-quark mass has been neglected with respect to the squark and gluino masses in the results of Ref.~\cite{Kauth:2011vg}. However, the numerical impact of this approximation is small. Applying the same translation procedure to the virtual part of the corrections to $q\bar{q}\to \tilde{g}\tilde{g}$, we find an agreement with the results presented in~\cite{Kauth:2011vg}, apart from the expression for $c_4(r)$ in Eq.~(32) of~\cite{Kauth:2011vg}, where the $\ln\Bigl(\frac{1+r}{2}\Bigr)$ term should read $\ln\Bigl(1+r\Bigr)$.\footnote{We thank P.~Marquard for confirming this typo in Eq.~(32) of~\cite{Kauth:2011vg}.}

\subsection{Numerical properties of the hard matching coefficients}

\begin{figure}[!h]
\hspace{-0.55cm}
\begin{tabular}{lll}
(a)\hspace{-0.3cm}\epsfig{file=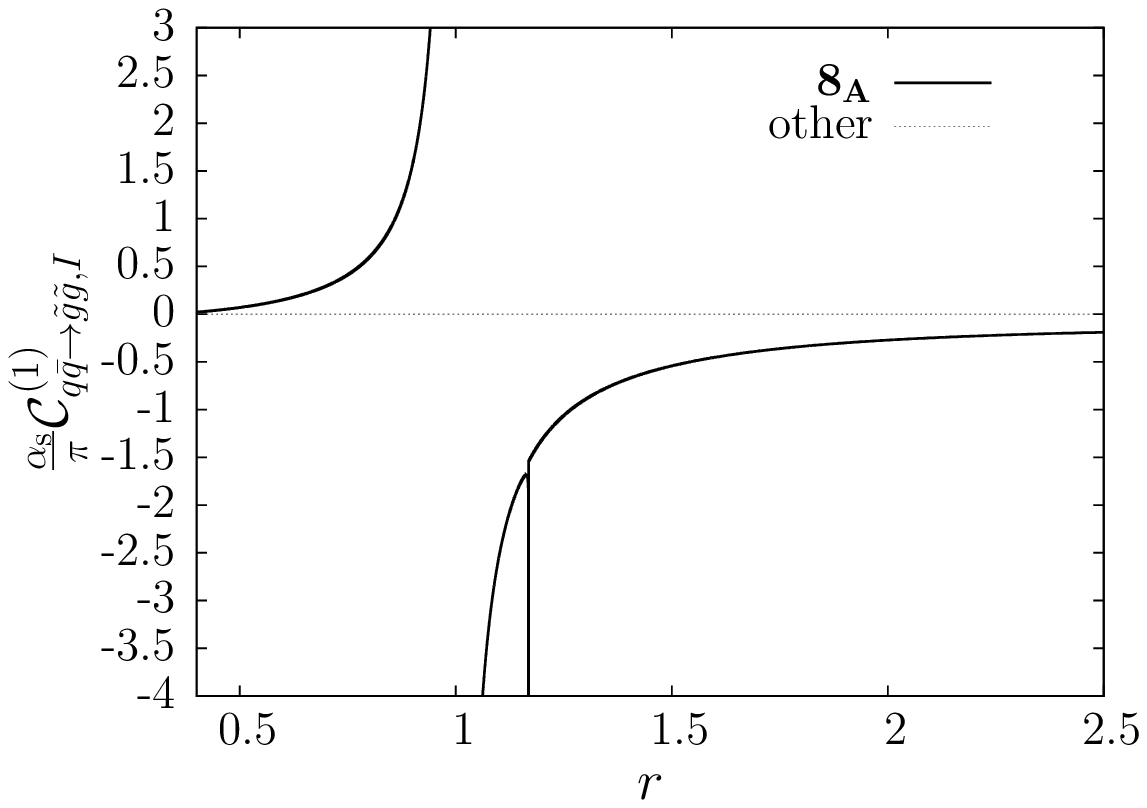, width=0.47\columnwidth}&
(b)\hspace{-0.3cm}\epsfig{file=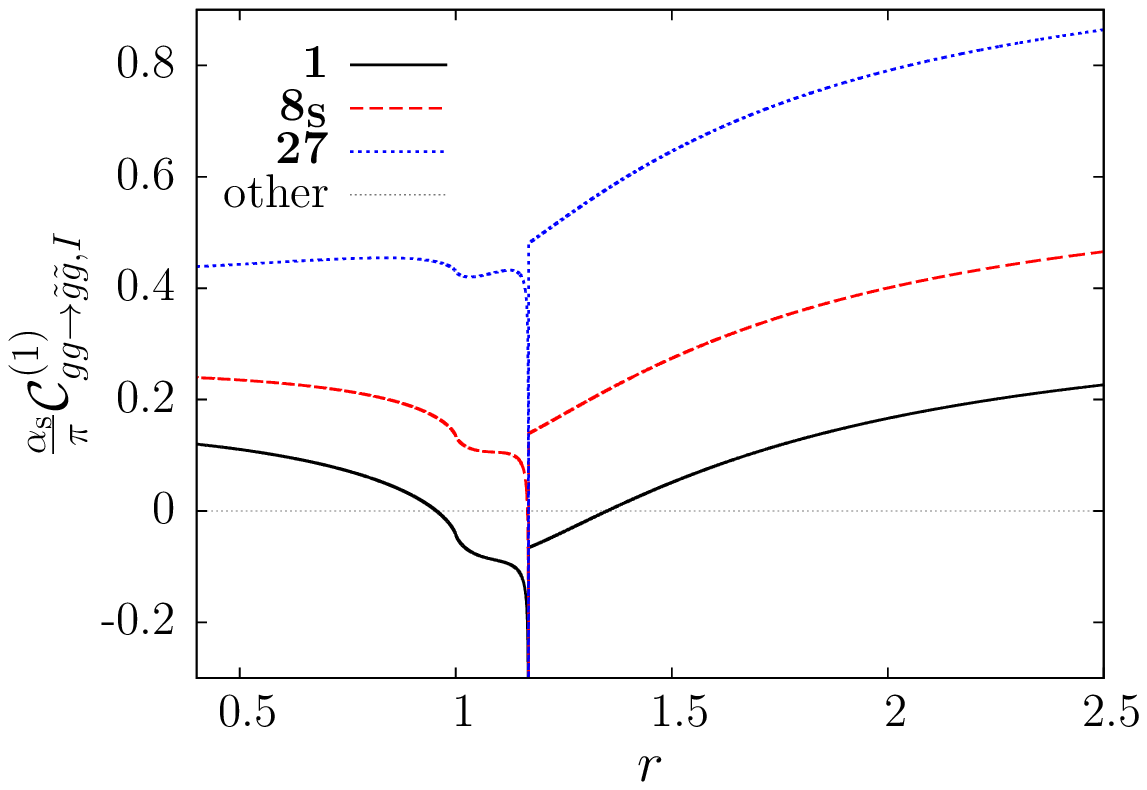, width=0.47\columnwidth}\\
(c)\hspace{-0.3cm}\epsfig{file=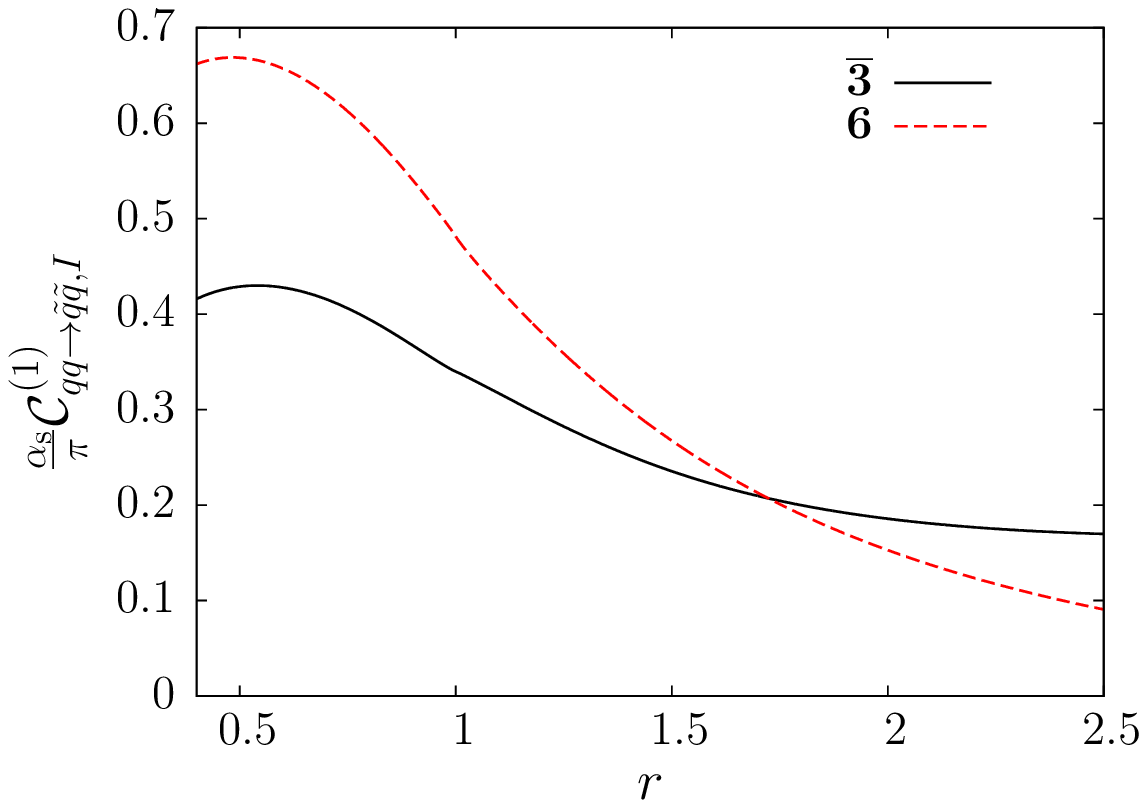, width=0.47\columnwidth}&
(d)\hspace{-0.3cm}\epsfig{file=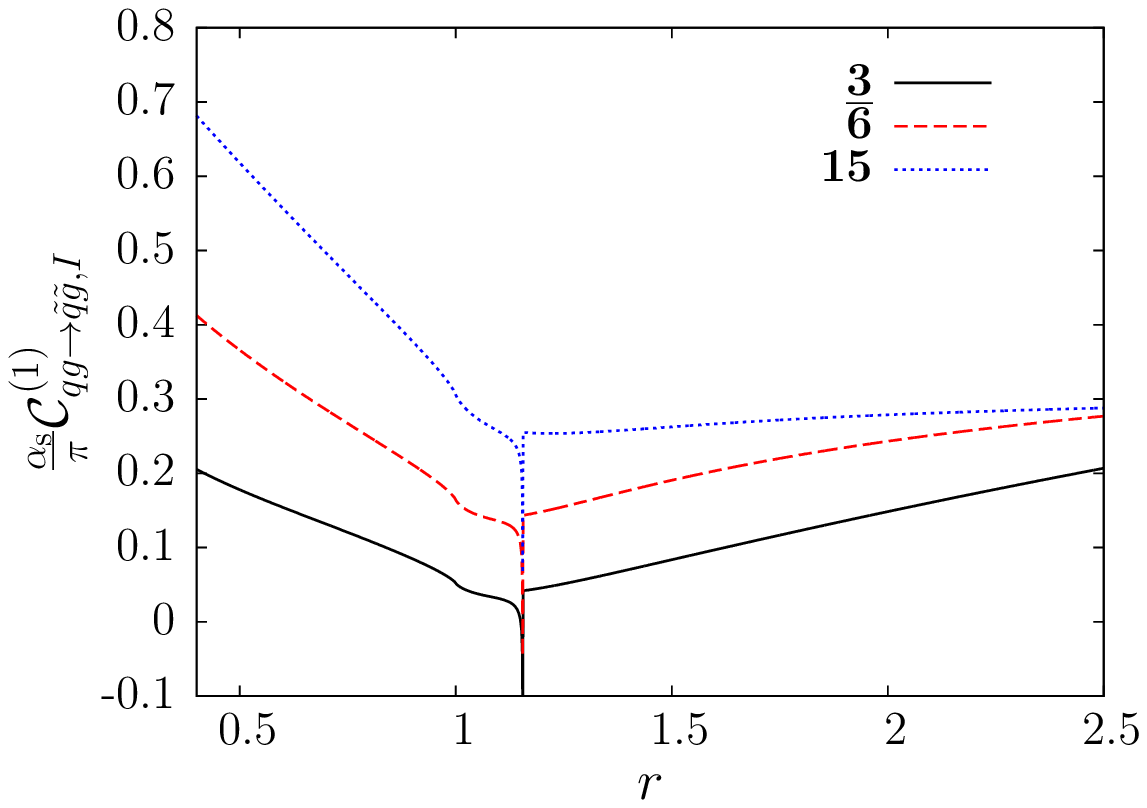, width=0.47\columnwidth}
\end{tabular}
\caption{Mass dependence of the SU(3) colour-decomposed NLO hard matching coefficients for the $q\bar q\to\gl\gl$ process (a), the $gg\to\gl\gl$ process (b), the $qq\to\sq\sq$ process (c) and the $qg\to\sq\gl$ process (d). The common renormalization and factorization scales have been set equal to the average mass of the produced particles $m_{av}=1.2$~TeV. The top quark mass is taken to be $m_t=172.9$~GeV.\label{fig:Ccoeffplots}}
\end{figure}

The behaviour of the hard matching coefficients for a varying squark-gluino mass ratio $r=m_\gl/m_\sq$ is shown in Fig.~\ref{fig:Ccoeffplots}. The corresponding plots for squark-antisquark production can be found in Ref.~\cite{Beenakker:2011sf}.

We see from Fig.~\ref{fig:Ccoeffplots} that the hard matching coefficients can be quite large depending on the masses involved. The processes that involve gluinos have a singularity if the gluino mass equals the sum of the squark and the top mass. This singularity could be removed by taking the finite gluino width into account. These matching coefficients have been checked numerically using {\tt PROSPINO}~\cite{Beenakker:1996ch}. As explained in section~\ref{s:zero}, the hard matching coefficients of suppressed cross sections have been set to 0.

As we can see from Fig.~\ref{fig:Ccoeffplots}(a), the hard matching coefficient ${\cal C}_{q\bar q\to\gl\gl,\bf 8_A}^{\rm (1)}$ for the $q\bar q\to\gl\gl$ process is ill-defined at $r=1$. This behaviour originates from the threshold behaviour of the LO cross section around $r=1$. For the \mbox{$q\bar q\to\gl\gl$} process, the leading term in the large $N$ expansion is proportional to $(m_\gl^2-m_\sq^2)^2$ at LO and thus vanishes accidentally for $m_\sq=m_\gl$. The first terms in the large $N$ expansion for the non-suppressed LO cross section in Mellin moment space are given by:
\begin{align}
\tilde\sigma_{q\bar q\to\gl\gl,\bf 8_A}^{\rm LO}(N)\approx\frac{\alpha_{\rm s}^2\pi^{3/2} (N_c^2-1) }{16 m_\gl^2 N_c N^{3/2}}\Bigg(B^2\!&-\!\frac{4\!+\!B\!-\!B^2}{N\!+\!\tfrac{3}{2}}B^2\!\label{eq:sigmaqqbtoggexp}\\
&+\!\frac{1\!+\!2B\!+\!39B^2\!+\!34B^3\!-\!12B^4\!-\!4B^5\!+\!9B^6}{4(N\!+\!\tfrac{3}{2})(N\!+\!\tfrac{5}{2})}\Bigg)\nonumber
\end{align}
with $B=\tfrac{r^2-1}{r^2+1}$, which vanishes for $r=1$. Thus the first two terms vanish for equal squark and gluino masses and the first non-zero term in the LO cross section is the ${\cal O}(N^{-7/2})$ term, which corresponds to a ${\cal O}(\beta^5)$ threshold behaviour. There is no symmetry that causes this behaviour, so it is not surprising that it does not hold at higher orders. In fact, the NLO threshold cross section contains terms proportional to $m_\gl^2-m_\sq^2$. To obtain the hard matching coefficient from the NLO threshold cross section, we have to divide the NLO cross section by the ${\cal O}(N^{-3/2})$ leading term in Eq.~\eqref{eq:sigmaqqbtoggexp}, resulting in a $(m_\gl^2-m_\sq^2)^{-1}$ divergence in the hard matching coefficient.

\begin{figure}[!h]
\hspace{-0.55cm}
\begin{tabular}{lll}
(a)\hspace{-0.3cm}\epsfig{file=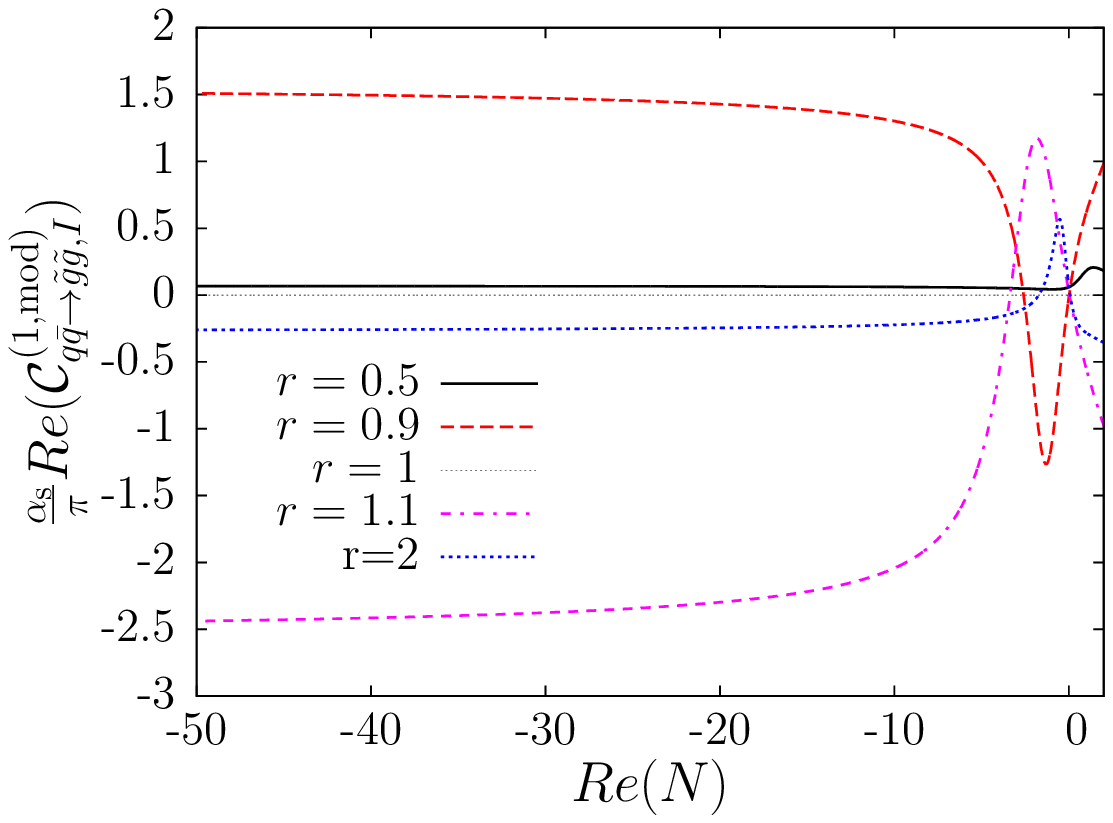, width=0.47\columnwidth}&
(b)\hspace{-0.3cm}\epsfig{file=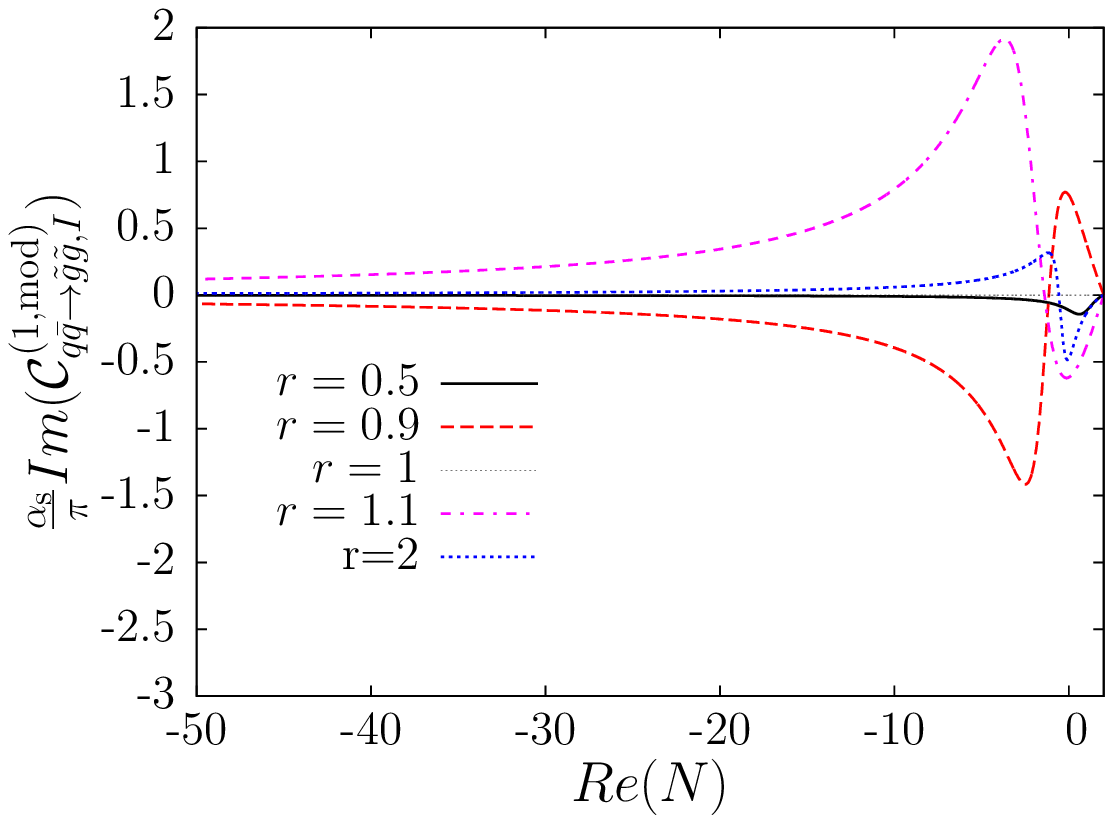, width=0.47\columnwidth}
\end{tabular}
\caption{Real and imaginary part of the modified non-vanishing $q\bar q\to\gl\gl$ hard matching coefficient in the $I=8A$ colour channel for different values of $r$ and $N=2+\exp{(i 3 \pi/4) \zeta},\ \zeta=0..\infty$. The common renormalization and factorization scale has been set equal to the average mass of the produced particles $m_{av}=1.2$~TeV. The top quark mass is taken to be $m_t=172.9$~GeV.\label{fig:CcoeffN}}
\end{figure}

Although the hard matching coefficient is multiplied by a cross section that is suppressed near $r=1$, such behaviour leads to numerical instabilities. For a stable numerical implementation, we thus propose to define a modified matching coefficient, which uses higher order terms in $1/N$ to regularize the divergence. Instead of using only the leading term in the large $N$ expansion, we use all three terms given in Eq.~\eqref{eq:sigmaqqbtoggexp}. After working out the Mellin transforms, the modified hard matching coefficient takes the form:
\begin{align}
{\cal C}_{q\bar q\to\gl\gl,{\bf 8A}}^{\rm(1,mod)}&=\bigg[1\!-\!\frac{4\!+\!B\!-\!B^2}{N\!+\!\tfrac{3}{2}}\!+\!\frac{1\!+\!2B\!+\!39B^2\!+\!34B^3\!-\!12B^4\!-\!4B^5\!+\!9B^6}{4B^2(N\!+\!\tfrac{3}{2})(N\!+\!\tfrac{5}{2})}\bigg]^{-1}{\cal C}_{q\bar q\to\gl\gl,{\bf 8A}}^{\rm(1)}\,,
\label{modC}
\end{align}
This deeper expansion of the LO cross section ensures a well-behaved hard matching coefficient, which vanishes at $r=1$. The hard matching coefficient now depends on $N$, which is a complex variable. In order to obtain the full cross section, we integrate over an appropriate contour in $N$-space, for which we use the ``minimal prescription'' of Ref.~\cite{Catani:1996yz}. The behaviour of both the real and the imaginary part of the hard matching coefficient is shown in Fig.~\ref{fig:CcoeffN}.  For large values of $|Re(N)|$, the modified hard matching coefficient approaches the value of the $N$-independent unmodified matching coefficient, cf. Eq.~(\ref{modC}), and the curves get flatter with increasing $|Re(N)|$ in Fig.~\ref{fig:CcoeffN}. As $r$ approaches 1, the second term in Eq.~(\ref{modC}) eventually takes over and ${\cal C}_{q\bar q\to\gl\gl,{\bf 8A}}^{\rm(1,mod)}$  becomes 0 (not shown in the plots), also for large (fixed) $|Re(N)|$.

\section{Numerical impact on LHC cross sections}
\label{s:numres}

In this section we illustrate the numerical impact of the hard matching coefficients on the NLL-resummed cross sections. For this we first have to perform the inverse Mellin transform in order to recover the hadronic cross section $\sigma_{h_1 h_2 \to kl}$. In order to retain the information contained in the complete NLO cross sections~\cite{Beenakker:1996ch}, the NLO and resummed results are combined through a matching procedure that avoids double counting of the NLO terms:
\begin{align}
  \label{eq:matching}
  &\sigma^{\rm (NLO+NLL+C^{\rm(1)}~matched)}_{h_1 h_2 \to kl}\bigl(\rho, \{m^2\},\mu^2\bigr) 
  =\; \sigma^{\rm (NLO)}_{h_1 h_2 \to kl}\bigl(\rho, \{m^2\},\mu^2\bigr)
          \\[1mm]
& \qquad+\, \sum_{i,j}\,\int_\mathrm{CT}\,\frac{dN}{2\pi i}\,\rho^{-N}\,
       \tilde f_{i/h_1}(N+1,\mu^2)\,\tilde f_{j/h_{2}}(N+1,\mu^2) \nonumber\\[2mm]
&\qquad\times\,
       \left[\tilde\sigma^{\rm(res,NLL+C^{\rm(1)})}_{ij\to kl}\bigl(N,\{m^2\},\mu^2\bigr)
             \,-\, \tilde\sigma^{\rm(res,NLL+C^{\rm(1)})}_{ij\to kl}\bigl(N,\{m^2\},\mu^2\bigr)
       {\left.\right|}_{\scriptscriptstyle{\rm (NLO)}}\, \right]. \nonumber
\end{align}
Here $\tilde f_{i/h_1}$ and $\tilde f_{j/h_2}$ are the Mellin transforms of the parton distribution functions. We adopt the ``minimal prescription'' of Ref.~\cite{Catani:1996yz} for
the contour CT of the inverse Mellin transform in Eq.~(\ref{eq:matching}).

We show the results for the LHC for a centre-of-mass energy of 8~TeV. In order to evaluate hadronic cross sections we use the 2008 NLO MSTW parton distribution functions~\cite{Martin:2009iq} with the corresponding $\alpha_{\rm s}(M_{Z}^2) = 0.120$. We have used a top quark mass of $m_t=172.9$~GeV~\cite{Nakamura:2010zzi}. The numerical results have been obtained with two independent computer codes. In order to evaluate resummed cross sections, we use both the method of Ref.~\cite{Kulesza:2002rh} with standard parametrization of pdfs in $x$-space, as well as the method first introduced in~\cite{Beenakker:2010nq}, relying on Mellin-space pdf obtained with the program {\tt PEGASUS}~\cite{Vogt:2004ns}. For comparison, we will also show results for the NLL matched cross section $\sigma^{\rm NLO+NLL}$ without the effect of the hard matching coefficient, as well as LO and NLO results. 

The NLO cross sections are calculated using the publicly available {\tt PROSPINO} code~\cite{prospino}, based on the calculations presented in Ref.~\cite{Beenakker:1996ch}.  The QCD coupling $\alpha_{\rm s}$ and the parton distribution functions at NLO are defined in the $\overline{\rm MS}$ scheme with five active flavours. The masses of squarks and gluinos are renormalized in the on-shell scheme, and both the top quark and the SUSY particles are decoupled from the running of $\alpha_{\rm s}$ and the parton distribution functions. No top-squark final states are considered.  We sum over squarks with both chiralities ($\tilde{q}_{L}$ and~$\tilde{q}_{R}$), which are taken as mass degenerate. The renormalization and factorization scales $\mu$ are taken to be equal.

\begin{figure}[!h]
\hspace{-0.55cm}
\begin{tabular}{lll}
(a)\hspace{-0.3cm}\epsfig{file=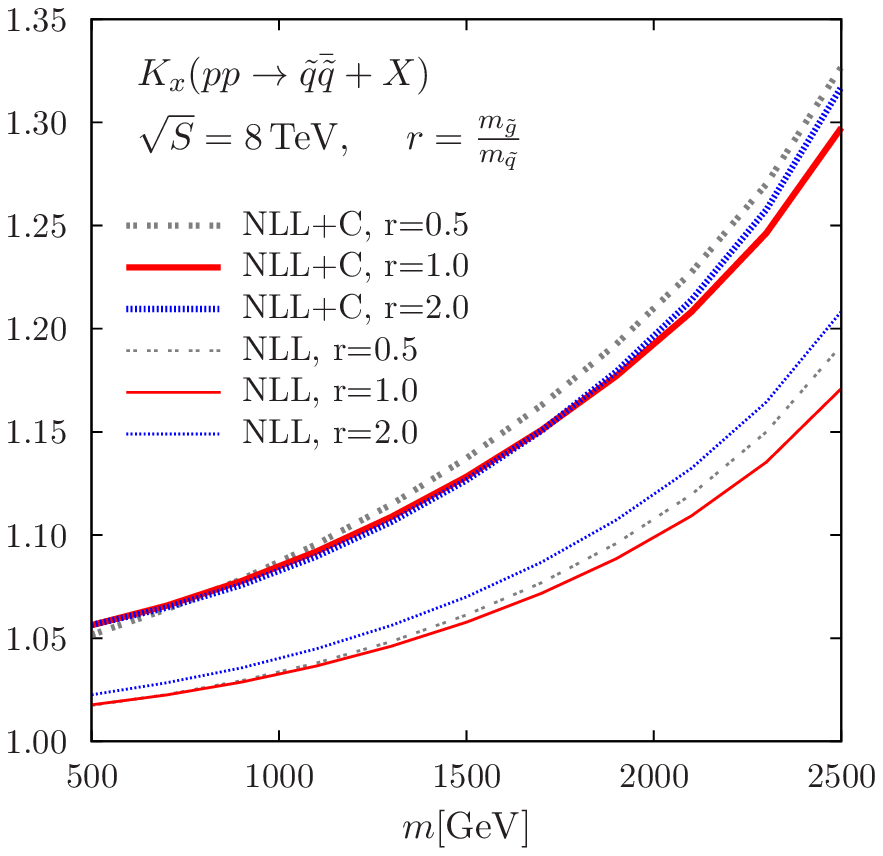, width=0.47\columnwidth}&
(b)\hspace{-0.3cm}\epsfig{file=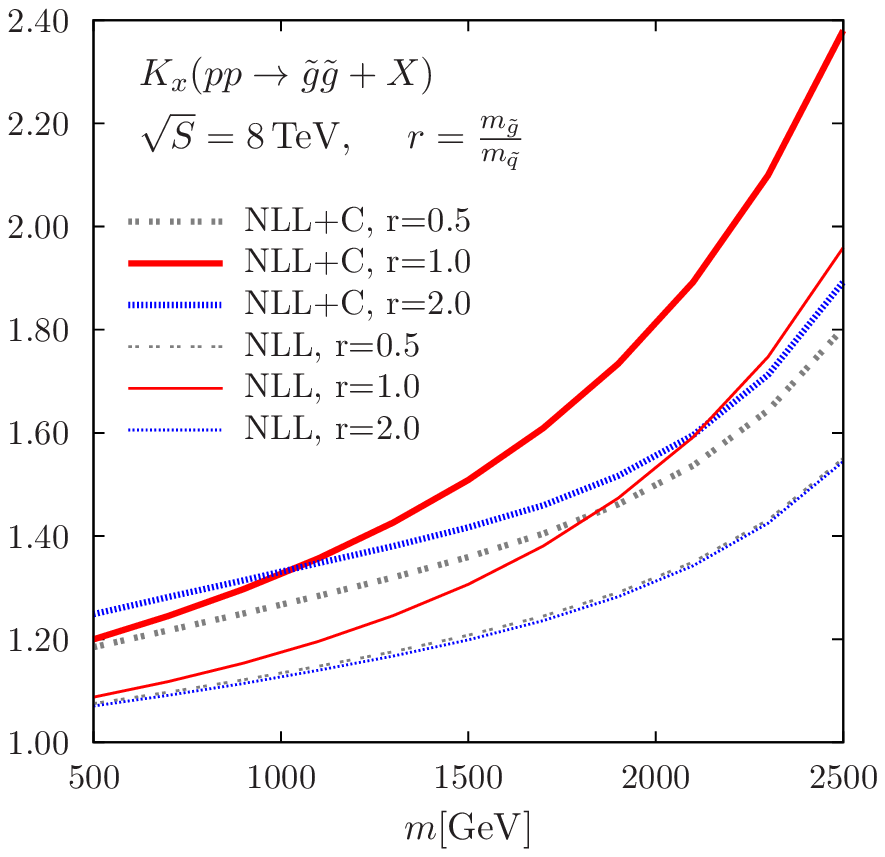, width=0.47\columnwidth}\\
(c)\hspace{-0.3cm}\epsfig{file=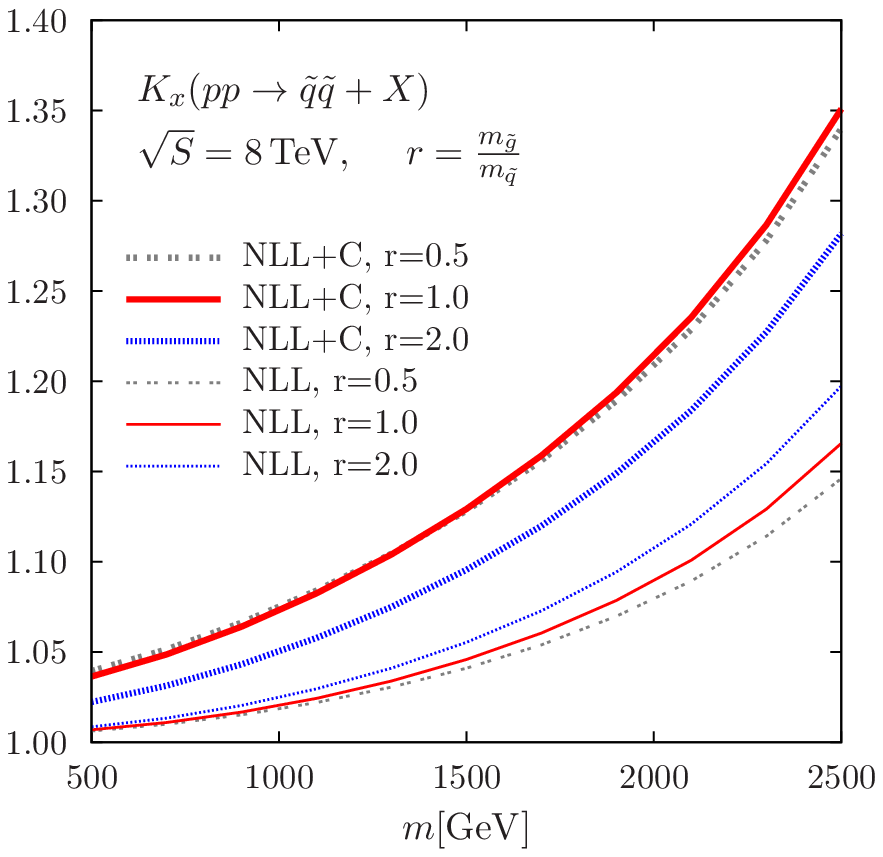, width=0.47\columnwidth}&
(d)\hspace{-0.3cm}\epsfig{file=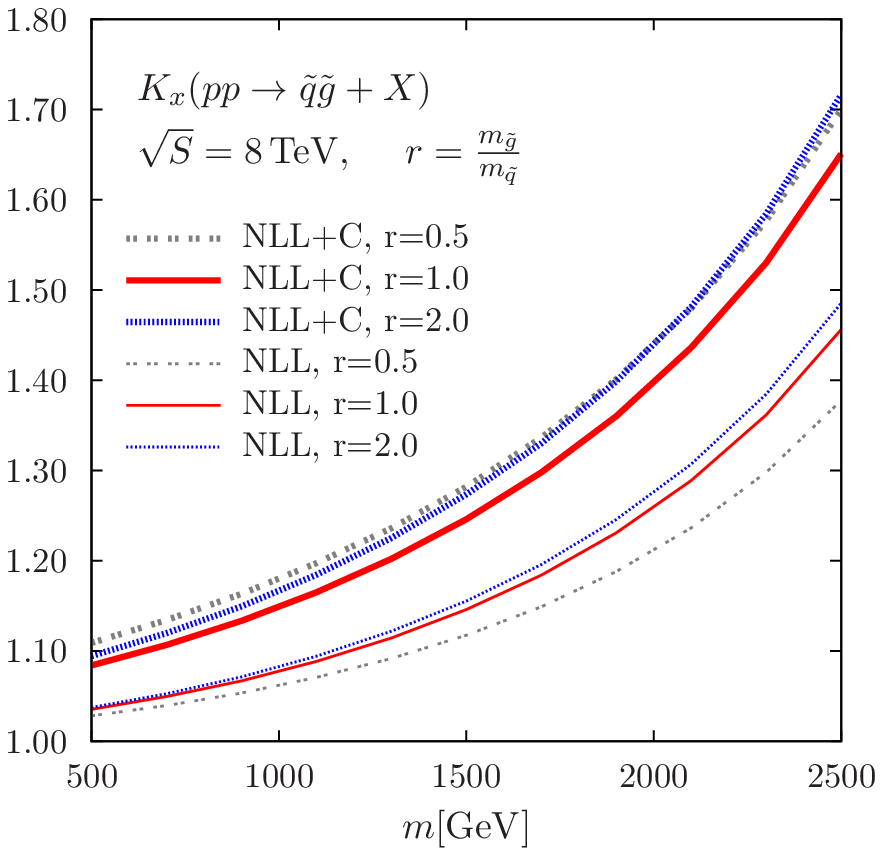, width=0.47\columnwidth}
\end{tabular}
\caption{The $K$-factor with respect to the NLO cross section of the NLO+NLL and NLO+NLL+C cross sections for the different SUSY-QCD processes for the LHC at 8 TeV. The common renormalization and factorization scales have been set equal to the average mass of the pair-produced particles $m=m_{av}$.\label{fig:K8}}
\end{figure}

We define the $K$-factors for the NLL matched cross section with and without the hard matching coefficient in the following way:
\[K_x=\frac{\sigma^x}{\sigma^{\rm (NLO)}}\ .\]
These $K$-factors are shown in Figure~\ref{fig:K8} for different mass ratios $r=m_\gl/m_\sq$.
We see that the hard matching coefficients give a correction to the cross section that is comparable to the NLL correction. The corrections range from a few percent to up to 10\% in the low-mass case, whereas they can reach several tens of percent in the high-mass case. Again the largest effects are observed for those final states that give rise to the largest colour factors. In particular for processes involving gluinos, it is vital to use resummed results for obtaining an accurate prediction of the cross section, with total resummation corrections that can exceed 100\% of the NLO cross section. The results are in general mildly mass-dependent, although a different mass ratio can lead to significant changes for gluino-pair production, and to a lesser extent for squark-pair production. The changes in the observed dependence on the mass ratio follow the $r$-dependence of the hard matching coefficients shown in Fig.~\ref{fig:Ccoeffplots}.

Although a full NNLL analysis is needed to draw definite conclusions on the size of the cross section and the scale dependence, these results indicate that the impact of the hard matching coefficients is quite significant. Thus we expect a fully NNLL-resummed cross section will improve the current theoretical predictions considerably.

\section{Conclusions}
\label{s:conclusion}

In this paper, we have completed the calculation of the hard matching coefficients for the squark and gluino hadroproduction processes. The hard matching coefficients are an important ingredient for NNLL resummation. Numerically, we find that the inclusion of the hard matching coefficients increases the cross section at the central scale.

\section*{Acknowledgments}

This work has been supported in part by the Helmholtz Alliance
``Physics at the Terascale'', the Foundation for Fundamental
Research of Matter (FOM), program 104 ``Theoretical Particle Physics in
the Era of the LHC", the DFG SFB/TR9 ``Computational Particle Physics'', Polish National Science Centre grant, project number DEC-2011/01/B/ST2/03643, European Community's Marie-Curie Research Training Network
under contract MRTN-CT-2006-035505 ``Tools and Precision Calculations
for Physics Discoveries at Colliders'' and by the Research Executive Agency (REA) of the European Union under the Grant Agreement number PITN-GA-2010-264564 (LHCPhenoNet).

\appendix
\allowdisplaybreaks
\section{Base tensors for SUSY-QCD}\label{app:colourstructures}

In this appendix, all the base tensors needed for the $2\to2$ SUSY-QCD processes are listed.  They are obtained with the method described in section~\ref{s:colour} and are given for a general SU$(N_c)$ theory in terms of Kronecker deltas in colour space $\delta_{ab}$, the generators of the fundamental representation $T^c_{ab}$, the structure constants $f_{abc}$ and their symmetric counterparts~$d_{abc}$. We give the dimension and quadratic Casimir invariant of all the base tensors. We use colour labels $a_1$ and $a_2$ for the initial-state particles and labels $a_3$ and~$a_4$ for the final-state particles. Summation over repeated indices is implied.\\

For the process $q(a_1)\bar q(a_2)\to\sq(a_3)\sqb(a_4)$ we have:
\begin{align}
c_{q\bar q\to\sq\sqb,1}&=\frac{1}{N_c}\delta_{a_1a_2}\delta_{a_3a_4}\,,&\dim(R_1)&=1\,,& C_2(R_1)&=0\,,\label{eq:qqbtosbc1}\\
c_{q\bar q\to\sq\sqb,2}&=2\,T^c_{a_2a_1}T^c_{a_3a_4}\,,&\dim(R_2)&=N_c^2-1\,,&C_2(R_2)&=N_c\,.\label{eq:qqbtosbc2}
\end{align}
In SU(3), we have $R_1={\bf 1}$ and $R_2={\bf 8}$. \\

For the process $g(a_1)g(a_2)\to\sq(a_3)\sqb(a_4)$ there are three structures:
\begin{align}
c_{gg\to\sq\sqb,1}&=\frac{1}{\sqrt{N_c(N_c^2-1)}}\delta_{a_1a_2}\delta_{a_3a_4}\,,&\dim(R_1)&=1\,,& C_2(R_1)&=0\,,\label{eq:ggtosbc1}\\
c_{gg\to\sq\sqb,2}&=\frac{i\sqrt{2}}{\sqrt{N_c}}f_{a_1a_2c}T^c_{a_3a_4}\,,&\dim(R_2)&=N_c^2-1\,,& C_2(R_2)&=N_c\,,\label{eq:ggtosbc2}\\
c_{gg\to\sq\sqb,3}&=\frac{\sqrt{2N_c}}{\sqrt{N_c^2-4}}d_{a_1a_2c}T^c_{a_3a_4}\,,&\dim(R_3)&=N_c^2-1\,,& C_2(R_3)&=N_c\,\label{eq:ggtosbc3}
\end{align}
where $R_1={\bf 1}$, $R_2={\bf 8_A}$ and $R_3={\bf 8_S}$ in SU(3). \\

For the squark-pair production process $q(a_1)q(a_2)\to\sq(a_3)\sq(a_4)$ the base tensors are:
\begin{align}
c_{qq\to\sq\sq,1}&=\frac{1}{2}\left(\delta_{a_1a_4}\delta_{a_2a_3}-\delta_{a_1a_3}\delta_{a_2a_4}\right)\,,&
c_{qq\to\sq\sq,2}&=\frac{1}{2}\left(\delta_{a_1a_4}\delta_{a_2a_3}+\delta_{a_1a_3}\delta_{a_2a_4}\right)\,,\label{eq:qqtoqqc}
\end{align}
and their dimension and quadratic Casimir invariants are given by:
\begin{align}
\dim(R_1)&=\frac{1}{2}N_c(N_c-1)\,,& C_2(R_1)=\frac{(N_c+1)(N_c-2)}{N_c}\,,\label{eq:qqtoqqcas1}\\
\dim(R_2)&=\frac{1}{2}N_c(N_c+1)\,,& C_2(R_2)=\frac{(N_c-1)(N_c+2)}{N_c}\,.\label{eq:qqtoqqcas2}
\end{align}
In SU$(3)$, $R_1=\overline{\bf 3}$, while $R_2={\bf 6}$.\\

The colour structures of the $q(a_1)\bar q(a_2)\to\gl(a_3)\gl(a_4)$ process are similar to Eqs.~(\ref{eq:ggtosbc1}-\ref{eq:ggtosbc3}):
\begin{align}
c_{q\bar q\to\gl\gl,1}&=\frac{1}{\sqrt{N_c(N_c^2-1)}}\delta_{a_1a_2}\delta_{a_3a_4}\,,&\dim(R_1)&=1\,,& C_2(R_1)&=0\,,\label{eq:qqbtoggc1}\\
c_{q\bar q\to\gl\gl,2}&=\frac{i\sqrt{2}}{\sqrt{N_c}}f_{a_3a_4c}T^c_{a_2a_1}\,,&\dim(R_2)&=N_c^2-1\,,& C_2(R_2)&=N_c\,,\label{eq:qqbtoggc2}\\
c_{q\bar q\to\gl\gl,3}&=\frac{\sqrt{2N_c}}{\sqrt{N_c^2-4}}d_{a_3a_4c}T^c_{a_2a_1}\,,&\dim(R_3)&=N_c^2-1,& C_2(R_3)&=N_c\,.\label{eq:qqbtoggc3}
\end{align}
In SU(3), we have $R_1={\bf 1}$, $R_2={\bf 8_A}$ and $R_3={\bf 8_S}$. \\

For the $g(a_1)g(a_2)\to\gl(a_3)\gl(a_4)$ process, the base tensors are given by:
\begin{align}
c_{gg\to\gl\gl,1}&=\frac{1}{N_c^2-1}\delta_{a_1a_2}\delta_{a_3a_4}\,,\label{eq:ggtoggc1}\\
c_{gg\to\gl\gl,2}&=\frac{N_c}{N_c^2-4}d_{a_1a_2c}d_{ca_3a_4}\,,\label{eq:ggtoggc2}\\
c_{gg\to\gl\gl,3}&=\frac{1}{N_c}f_{a_1a_2c}f_{ca_3a_4}\,,\label{eq:ggtoggc3}\\
c_{gg\to\gl\gl,4}&=\frac{1}{4}\left(\delta_{a_1a_3}\delta_{a_2a_4}\!-\delta_{a_1a_4}\delta_{a_2a_3}\right)-\frac{f_{a_1a_2c}f_{ca_3a_4}}{2N_c}+\frac{i}{4}\left(d_{a_1a_3c}f_{ca_2a_4}\!+\!f_{a_1a_3c}d_{ca_2a_4}\right),\label{eq:ggtoggc4}\\
c_{gg\to\gl\gl,5}&=\frac{1}{4}\left(\delta_{a_1a_3}\delta_{a_2a_4}\!-\delta_{a_1a_4}\delta_{a_2a_3}\right)-\frac{f_{a_1a_2c}f_{ca_3a_4}}{2N_c}-\frac{i}{4}\left(d_{a_1a_3c}f_{ca_2a_4}\!+\!f_{a_1a_3c}d_{ca_2a_4}\right),\label{eq:ggtoggc5}\\
c_{gg\to\gl\gl,6}&=-\frac{N_c+2}{2N_c(N_c+1)}\delta_{a_1a_2}\delta_{a_3a_4}+\frac{N_c+2}{4N_c}\left(\delta_{a_1a_3}\delta_{a_2a_4}+\delta_{a_1a_4}\delta_{a_2a_3}\right)\label{eq:ggtoggc6}\\
&\qquad\qquad-\frac{N_c+4}{4(N_c+2)}d_{a_1a_2c}d_{a_3a_4c}+\frac{1}{4}\left(d_{a_1a_3c}d_{a_2a_4c}+d_{a_2a_3c}d_{a_1a_4c}\right)\,,\nonumber\\
c_{gg\to\gl\gl,7}&=\frac{N_c-2}{2N_c(N_c-1)}\delta_{a_1a_2}\delta_{a_3a_4}+\frac{N_c-2}{4N_c}\left(\delta_{a_1a_3}\delta_{a_2a_4}+\delta_{a_1a_4}\delta_{a_2a_3}\right)\label{eq:ggtoggc7}\\
&\qquad\qquad+\frac{N_c-4}{4(N_c-2)}d_{a_1a_2c}d_{a_3a_4c}\nonumber-\frac{1}{4}\left(d_{a_1a_3c}d_{a_2a_4c}+d_{a_2a_3c}d_{a_1a_4c}\right)\,,
\end{align}
while the corresponding dimensions and quadratic Casimir invariants are:
\begin{align}
\dim(R_1)&=1\,,&\quad&C_2(R_1)=0\,,\\[1mm]
\dim(R_2)&=N_c^2-1\,,&\quad&C_2(R_2)=N_c\,,\\[1mm]
\dim(R_3)&=N_c^2-1\,,&\quad&C_2(R_3)=N_c\,,\\[1mm]
\dim(R_4)&=(N_c^2-4)(N_c^2-1)/4\,,&\quad&C_2(R_4)=2N_c\,,\\[1mm]
\dim(R_5)&=(N_c^2-4)(N_c^2-1)/4\,,&\quad&C_2(R_5)=2N_c\,,\\[1mm]
\dim(R_6)&=N_c^2(N_c+3)(N_c-1)/4\,,&\quad&C_2(R_6)=2(N_c+1)\,,\\[1mm]
\dim(R_7)&=N_c^2(N_c-3)(N_c+1)/4\,,&\quad&C_2(R_7)=2(N_c-1)\,.
\end{align}
Note that since the dimension of $R_7$ vanishes for $N_c=3$, this representation does not contribute in (SUSY)-QCD. The other representations correspond to $R_1={\bf 1}$, $R_2={\bf 8_S}$ and  $R_3={\bf 8_A}$. The $R_4$ and $R_5$ are the $\bf10$ and $\bf \overline{10}$ representations, while $R_6=\bf27$ in SU(3). \\

Finally, the base tensors for squark-gluino production $q(a_1)g(a_2)\to\sq(a_3)\gl(a_4)$ are given by:
\begin{align}
c_{qg\to\sq\gl,1} &=\frac{2N_c}{N_c^2-1}(T^{a_4}T^{a_2})_{a_3a_1}\,,\label{eq:qgtoqgc1}\\
c_{qg\to\sq\gl,2} &=\frac{N_c-2}{2N_c}\delta_{a_2a_4}\delta_{a_1a_3}-d_{ca_4a_2}T^c_{a_3a_1}
+\frac{N_c-2}{N_c-1}(T^{a_4}T^{a_2})_{a_3a_1}\,,\label{eq:qgtoqgc2}\\
c_{qg\to\sq\gl,2} &=\frac{N_c+2}{2N_c}\delta_{a_2a_4}\delta_{a_1a_3}+d_{ca_4a_2}T^c_{a_3a_1}
-\frac{N_c+2}{N_c+1}(T^{a_4}T^{a_2})_{a_3a_1}\,.\label{eq:qgtoqgc3}
\end{align}
The dimensions and quadratic Casimir invariants for the corresponding representations are:
\begin{align}
\dim(R_1)&=N_c\,,&C_2(R_1)&=\frac{N_c^2-1}{2N_c}\,,\\
\dim(R_2)&=\frac{1}{2}N_c(N_c+1)(N_c-2)\,,&C_2(R_2)&=\frac{(N_c-1)(3N_c+1)}{2N_c}\,,\\
\dim(R_3)&=\frac{1}{2}N_c(N_c-1)(N_c+2)\,,&C_2(R_3)&=\frac{(N_c+1)(3N_c-1)}{2N_c}\,,
\end{align}
which correspond to $R_1=\bf 3$, $R_2=\bf \overline{6}$ and $R_3=\bf15$ in SU(3).

\section{Hard matching coefficients for SUSY-QCD}\label{app:Ccoeff}

Here we present the exact expressions for the hard matching coefficients ${\cal C}^{\rm(1)}$ for the SUSY-QCD production processes. We sum over squarks with both chiralities ($\tilde{q}_{L}$ and $\tilde{q}_{R}$). No top-squark final states are considered and all squarks are considered to be mass-degenerate with mass $m_\sq$. Top squarks are taken into account in the loops, where they are taken to be mass-degenerate with the other squarks. This is a valid approximation, since SUSY parameter dependence coming from loop corrections is generally small~\cite{Beenakker:1996ch}. The calculation is outlined in section~\ref{s:Ccoeff} and was done with FORM \cite{Vermaseren:2000nd}. We first define the functions:
\[\beta_{12}(q^2)=\sqrt{1 - \frac{4m_1m_2}{q^2 - (m_1 - m_2)^2}}~,\qquad x_{12}(q^2)=\frac{\beta_{12}(q^2)-1}{\beta_{12}(q^2)+1}\]
and the combinations:
\[m_-^2=m_{\tilde g}^2-m_{\tilde q}^2,\qquad m_+^2=m_{\tilde g}^2+m_{\tilde q}^2,\]
where $m_\gl$ is the gluino mass and $m_\sq$ the squark mass. Furthermore, $m_{av}$ is the average mass of the produced particles. Denoting the number of light flavours by $n_l=5$, the total number of flavours by $n_f=6$ and the number of colours by $N_c$, we also define:
\begin{align*}
\gamma_q&=\frac{3}{2}C_F&C_F&=\frac{N_c^2-1}{2N_c}\\
\gamma_g&=\frac{11}{6}C_A-\frac{1}{3}n_l&C_A&=N_c\,.
\end{align*}
We denote the factorization scale by $\mu_F$, the renormalization scale by $\mu_R$ and Euler's constant by $\gamma_E$. The dilogarithm is defined as 
\[\mathrm{Li}_2(z)=-\int_0^z\frac{\log(1-t)\,\d t}{t}\]
We split the hard matching coefficients into a representation-independent part and a part that depends on the irreducible representation in the colour decomposition of Appendix~\ref{app:colourstructures}. 

The nonzero hard matching coefficients for the \mbox{$q\bar q\to\tilde q\bar{\tilde q}$} and the \mbox{$qq\to\sq\sq$} process are the same provided that the appropriate representations are used. Also, as we saw in section~\ref{s:zero}, the contribution of the antisymmetric colour structure of the squark-squark cross section is suppressed for squarks that have the same flavour, so the corresponding hard matching coefficient is zero:
\begin{align*}
{\cal C}_{qq\to\sq\sq,1}^{(1)}&=0\mbox{ for equal flavours}\\
{\cal C}_{q\bar q\to\tilde q\bar{\tilde q},I}^{(1)}&={\cal C}_{qq\to\sq\sq,I}^{(1)}=\mathrm{Re}\Bigg\{\frac{2C_F}{3}\pi^2+\gamma_g\log\bigg(\frac{\mu_R^2}{m_{\tilde q}^2}\bigg)-\gamma_q\log\bigg(\frac{\mu_F^2}{m_{\tilde q}^2}\bigg)+\frac{19N_c}{24}+\frac{23}{8N_c}\\
&-\frac{2}{N_c}\log\left(2\right)+\bigg(\frac{7N_c}{6}+\frac{2m_{\tilde g}^2}{m_+^2}C_F\bigg)\log\bigg(\frac{m_{\tilde g}^2}{m_{\tilde q}^2}\bigg)-\frac{m_{\tilde g}^2}{m_{\tilde q}^2}\bigg[\frac{m_-^2}{m_{\tilde q}^2}\log\bigg(\frac{m_-^2}{m_{\tilde g}^2}\bigg)+1\bigg]C_F\\
&+F_0(m_{\tilde q},m_{\tilde g},m_t)-\frac{1}{2N_c}\bigg(\frac{m_{\tilde g}^2}{m_{\tilde q}^2}-3\bigg)\,F_1\left(m_{\tilde q},m_{\tilde g}\right)+\bigg(\frac{m_+^2}{2m_{\tilde q}^2}C_F+\frac{1}{N_c}\bigg)F_2\left(m_{\tilde q},m_{\tilde g}\right)\\
&+2C_F\bigg[\gamma_E^2 + \gamma_E\log\bigg(\frac{\mu_F^2}{4m_\sq^2}\bigg)\bigg]+\frac{m_{\tilde g}^2}{2m_-^2}\bigg[\frac{m_{\tilde g}^2}{m_-^2}\log\!\bigg(\!\frac{m_{\tilde g}^2}{m_{\tilde q}^2}\!\bigg)-1\bigg]C_F+\frac{1-3N_c^2}{N_c}\log\!\bigg(\!\frac{m_+^2}{m_{\tilde q}^2}\!\bigg)\\
&+\bigg\{-\frac{\pi^2}{4}+\log\bigg(\frac{m_+^2}{m_{\tilde q}^2}\bigg)-\log\left(2\right)-\frac{m_{\tilde g}^2}{m_+^2}\log\bigg(\frac{m_{\tilde g}^2}{m_{\tilde q}^2}\bigg)+2+\gamma_E\\
&-\frac{1}{4}\bigg(\frac{m_{\tilde g}^2}{m_{\tilde q}^2}-3\bigg)\left[F_1\left(m_{\tilde q},m_{\tilde g}\right)+F_2\left(m_{\tilde q},m_{\tilde g}\right)\right]\bigg\}\,C_2(R_I)\Bigg\}.
\end{align*}
In this equation the last two lines are proportional to the quadratic Casimir invariants of the representations, which are given in Eqs. \eqref{eq:qqbtosbc1} and \eqref{eq:qqbtosbc2} for the $q\bar q\to\sq\sqb$ process and in Eqs. \eqref{eq:qqtoqqcas1} and \eqref{eq:qqtoqqcas2} for the $qq\to\sq\sq$ process. Furthermore we have defined the functions:
\begin{align*}
F_0(m_{\tilde q},m_{\tilde g},m_t)&=\frac{m_t^2}{2m_{\tilde g}^2}-\bigg(1+\frac{m_{\tilde q}^2}{2m_{\tilde g}^2}\bigg)\,n_f+\bigg[\frac{m_-^6}{2m_+^2m_{\tilde g}^4}\log\bigg(\frac{m_-^2}{m_{\tilde q}^2}\bigg)+\frac{4m_{\tilde q}^2}{m_+^2}\log\left(2\right)\!\bigg]n_l\\
&\quad+\bigg[\frac{m_t^4}{2m_{\tilde q}^2m_{\tilde g}^2}-\frac{(m_{\tilde q}^2-m_t^2)^2}{4m_{\tilde g}^4}+\frac{m_{\tilde q}^2-m_t^2}{m_{\tilde g}^2}-\frac{1}{12}\bigg]\log\bigg(\frac{m_t^2}{m_{\tilde q}^2}\bigg)\\
&\quad-\frac{m_-^2\big(m_{\tilde g}^2-(m_{\tilde q}-m_t)^2\big)\big(m_{\tilde g}^2-m_{\tilde q}^2+m_t^2\big)}{2m_{\tilde g}^4m_+^2}\beta_{\tilde qt}(m_{\tilde g}^2)\log\left(x_{\tilde qt}(m_{\tilde g}^2)\right)\\
&\quad+\frac{m_t^4-2m_{\tilde q}m_t^3+4m_{\tilde q}^3m_t-4m_{\tilde q}^4}{m_{\tilde q}^2m_+^2}\beta_{\tilde qt}(-m_{\tilde q}^2)\log\left(x_{\tilde qt}(-m_{\tilde q}^2)\right)\\
F_1(m_{\tilde q},m_{\tilde g})&=\mathrm{Li}_2\bigg(\frac{m_-^2}{2m_{\tilde g}^2}\bigg)+\mathrm{Li}_2\bigg(1-\frac{m_-^2}{2m_{\tilde q}^2}\bigg)\\
&\quad+\log\bigg(\frac{m_-^2}{2m_{\tilde g}^2}\bigg)\log\bigg(\frac{m_+^2}{2m_{\tilde q}^2}\bigg)+\frac{1}{2}\log^2\!\bigg(\frac{m_{\tilde g}^2}{m_{\tilde q}^2}\bigg)+\frac{\pi^2}{12}\\
F_2(m_{\tilde q},m_{\tilde g})&=\mathrm{Li}_2\bigg(\frac{m_{\tilde q}^2}{m_{\tilde g}^2}\bigg)-\mathrm{Li}_2\bigg(-\frac{m_{\tilde q}^2}{m_{\tilde g}^2}\bigg)+\log\bigg(\frac{m_+^2}{m_-^2}\bigg)\log\bigg(\frac{m_{\tilde g}^2}{m_{\tilde q}^2}\bigg)\,.
\end{align*}

For the $gg\to\tilde q\bar{\tilde q}$ process the antisymmetric representation in Eq.~\eqref{eq:ggtosbc2} does not contribute because it yields a $p$-wave contribution, which is suppressed near threshold. The hard matching coefficients for the representations of Eqs.~\eqref{eq:ggtosbc1} and~\eqref{eq:ggtosbc3} do contribute:
\begin{align*}
{\cal C}_{gg\to\tilde q\bar{\tilde q},2}^{(1)}&=0\\
{\cal C}_{gg\to\tilde q\bar{\tilde q},I}^{(1)}&=\mathrm{Re}\Bigg\{\bigg(\frac{5N_c}{12}-\frac{C_F}{4}\bigg)\,\pi^2+\gamma_g\log\bigg(\frac{\mu_R^2}{\mu_F^2}\bigg)-\frac{m_{\tilde g}^2N_c}{2m_{\tilde q}^2}\log^2\left(x_{\tilde g\tilde g}(4m_{\tilde q}^2)\right)\\
&+C_F\bigg[\frac{m_+^2m_-^2}{2m_{\tilde q}^4}\log\bigg(\frac{m_+^2}{m_-^2}\bigg)-\frac{m_{\tilde g}^2}{m_{\tilde q}^2}-3\bigg]+\frac{m_+^2N_c}{2m_{\tilde q}^2}\bigg[\mathrm{Li}_2\bigg(\!-\!\frac{m_{\tilde q}^2}{m_{\tilde g}^2}\bigg)-\mathrm{Li}_2\bigg(\frac{m_{\tilde q}^2}{m_{\tilde g}^2}\bigg)\bigg]\\
&+2C_A\bigg[\gamma_E^2 + \gamma_E\log\bigg(\frac{\mu_F^2}{4m_\sq^2}\bigg)\bigg]\\
&+\bigg\{\frac{\pi^2}{8}-\frac{1}{2}\mathrm{Li}_2\bigg(\!\!-\!\frac{m_{\tilde q}^2}{m_{\tilde g}^2}\bigg)+\frac{1}{2}\mathrm{Li}_2\bigg(\frac{m_{\tilde q}^2}{m_{\tilde g}^2}\bigg)+\frac{m_{\tilde g}^2}{4m_{\tilde q}^2}\log^2\!\left(x_{\tilde g\tilde g}(4m_{\tilde q}^2)\right)+2+\gamma_E\bigg\}C_2(R_I)\Bigg\}\,,
\end{align*}
where in the second equation the representation $I$ can be either of the symmetric representations, with colour tensors given by Eq.~\eqref{eq:ggtosbc1} or~\eqref{eq:ggtosbc3}, and the last line is proportional to the corresponding quadratic Casimir invariant. 

In the case of the $q\bar q\to\gl\gl$ process, only the antisymmetric representation in Eq.~\eqref{eq:qqbtoggc2} yields a nonzero matching coefficient, the cross sections for the other representations are suppressed near threshold:
\begin{align*}
{\cal C}_{q\bar q\to\gl\gl,1}^{(1)}&={\cal C}_{q\bar q\to\gl\gl,3}^{(1)}=0\\
{\cal C}_{q\bar q\to\gl\gl,2}^{(1)}&=\mathrm{Re}\Bigg\{ \frac{m_\sq^4C_F}{4m_\gl^2m_-^2} + \frac{7}{4N_c} + \frac{3m_\sq^2}{8m_-^2N_c} + \frac{5(2m_\gl^2 + m_\sq^2)N_c}{24m_-^2}+\gamma_g\log\!\bigg(\!\frac{\mu_R^2}{m_\gl^2}\!\bigg)-\gamma_q\log\!\bigg(\!\frac{\mu_F^2}{m_\gl^2}\!\bigg)\\
&+\frac{N_c^2-4}{12N_c}\pi^2+2C_F \bigg[\gamma_E^2+\gamma_E\log\bigg(\frac{\mu_F^2}{4m_\gl^2}\bigg)\bigg]+ \Bigg[\frac{m_\gl^2m_\sq^2}{4m_-^4N_c}+\frac{3}{8N_c}+ \frac{m_\sq^2}{m_-^2N_c} + \frac{m_\sq^4C_F}{m_\gl^2m_-^2}\\
& - \frac{m_\sq^4m_+^2N_c}{8m_\gl^2m_-^4}+ \frac{m_\sq^2N_c}{8m_-^2} - \frac{2m_\sq^4N_c}{m_-^2m_+^2} \Bigg]\log\bigg(\frac{m_\sq^2}{m_\gl^2}\bigg)+ \frac{1 - 2 N_c^2}{N_c}\bigg[\frac{m_\gl^2}{m_-^2}+\frac{m_-^2}{4m_\gl^2}\bigg] \log\bigg(\frac{m_+^2}{m_\gl^2}\bigg)\\
&+ \Bigg[\frac{2m_\gl^2m_-^2C_F}{m_\sq^2m_+^2} -\frac{m_+^2}{4m_\gl^2N_c}-\frac{m_\gl^2}{m_-^2N_c}\Bigg]\log\bigg(\frac{m_-^2}{m_\gl^2}\bigg)+F_5(m_\sq,m_\gl,m_t)+\gamma_E N_c\\
 &+ \Bigg[ \bigg(5+\frac{m_+^2}{m_-^2} + \frac{2m_-^2}{m_+^2}\bigg)C_F - \frac{5N_c}{2}\Bigg]\log(2) - \frac{m_+^4 (5 m_\gl^4 - 2 m_\gl^2 m_\sq^2 + m_\sq^4)}{32 m_\gl^4 m_-^4 N_c}F_4(m_\gl,m_\sq) \\
 &- \frac{(3 m_\gl^2 - m_\sq^2)m_+^4 }{ 8 m_\gl^2 m_-^4 N_c}\log\big(x_{\sq\sq}(4m_\gl^2)\big)\,\beta_{\sq\sq}(4m_\gl^2)\\
 &+  \frac{1}{N_c}\Bigg[\frac{m_\gl^2(m_\sq^2 - 3m_\gl^2)}{2m_-^4} - \frac{m_-^2}{8m_\gl^2} - \frac{1}{4}\Bigg]F_1(m_\gl,m_\sq)\\
&+\frac{N_c}{8}\Bigg[\frac{m_-^4}{2m_\gl^4} - \frac{m_\sq^2}{m_\gl^2} - \frac{3(3m_\gl^2 + m_\sq^2)}{m_-^2}\Bigg] F_2(m_\gl,m_\sq)\Bigg\}\,,
\end{align*}
where we have defined the additional functions:
\begin{align*}
F_4(m_\gl, &m_\sq) = \mathrm{Li}_2\bigg( 1 - x_{\sq\sq}(4m_\gl^2)\frac{m_\gl^2}{m_\sq^2}\bigg) +\mathrm{Li}_2\bigg( 1 - \frac{1}{x_{\sq\sq}(4m_\gl^2)}\frac{m_\gl^2}{m_\sq^2}\bigg) \\
&\qquad\qquad-2\mathrm{Li}_2\bigg(1-\frac{ m_\gl^2}{m_\sq^2}\bigg) + \log^2\big(x_{\sq\sq}(4m_\gl^2)\big)\,,\\
F_5(m_\sq,&m_\gl,m_t)=\frac{(m_t^2 - m_\sq^2n_f)(2m_\gl^2 - m_\sq^2)}{m_\gl^2m_-^2}+  \Bigg[\frac{2m_\sq(3m_\gl^2+m_\sq^2)m_t^2(m_t-m_\sq)}{3m_\gl^2m_-^2(m_+^2-m_t^2)}+\frac{m_+^2m_t^2}{m_\gl^2m_-^2}\\
&-\frac{2m_\sq^3m_+^2m_t}{3m_\gl^4m_-^2}+\frac{(5m_\gl^2-4m_\sq^2)(2m_\sq-m_t)m_t^3}{3m_\gl^4m_-^2}+\frac{2m_+^2(m_\sq-m_t)m_t}{3m_\gl^2(m_-^2+m_t^2)}\\
&-\frac{2m_\sq m_-^2(2m_\sq-5m_t)}{3m_\gl^4}+\frac{4m_\sq m_t(m_\sq m_t-m_-^2)}{m_\gl^2(m_\gl^2-(m_\sq+m_t)^2)}\Bigg]\log\big(x_{\sq t}(m_\gl^2)\big)\beta_{\sq t}(m_\gl^2)\\
&+\Bigg[ \frac{ m_t^2}{2m_-^2} + \frac{2m_\sq^2 \big(m_\gl^4+(m_\sq^2 - m_t^2)^2\big)}{3m_\gl^4m_-^2} -\frac{8 m_\sq^4 - 13 m_\sq^2 m_t^2 + 5 m_t^4}{6 m_\gl^2 m_-^2}\Bigg]\log\bigg(\frac{m_t^2}{m_\sq^2}\bigg)\\
&+ \frac{m_+^2}{m_-^2}\Bigg[ \frac{m_t^2}{3(m_-^2 + m_t^2)} -\frac{m_+^2n_f}{6m_\gl^2} \Bigg]\log\big(x_{\sq\sq}(4m_\gl^2)\big)\,\beta_{\sq\sq}(4m_\gl^2)+\frac{2n_l}{3}\log(2) \\
&+ \frac{m_+^2}{3m_-^2} \Bigg[ 2 +\frac{m_t^2}{2m_\gl^2} - \frac{3 m_\gl^2 + m_\sq^2}{ m_+^2- m_t^2}\Bigg]\log\big(x_{tt}(4m_\gl^2)\big)\,\beta_{tt}(4m_\gl^2)+\frac{4m_\sq^2m_-^2n_l}{3m_\gl^4}\log\bigg(\frac{m_-^2}{m_\sq^2}\bigg)\,.
\end{align*}

For the hard matching coefficients of the $gg\to\gl\gl$ process, the representation-dependent part does not scale with the quadratic Casimir invariants from Eq.~\eqref{eq:qqbtoggc3} due to contributions from box diagrams. Therefore we introduce the additional colour factors $C'(R_I)$ for convenience:
\[C'(R_1)=\frac{C_F}{2N_c}\ ,\qquad C'(R_2)=\frac{N_c^2-4}{4N_c^2}\ ,\qquad C'(R_6)=C'(R_7)=0\ .\]
We also introduce the function:
\begin{align*}
F_3(q_1^2, q_2^2, m_\sq, m_t)&= \log^2\bigg(x_{\sq t}(q_2^2)\frac{m_\sq}{m_t}\bigg) - \log^2\bigg(x_{\sq t}(q_1^2)\frac{m_\sq}{m_t}\bigg) - 2\mathrm{Li}_2\bigg( 1 - x_{\sq t}(q_2^2)\frac{m_t}{m_\sq}\bigg)\\
&\!\!\!\! + 2\mathrm{Li}_2\bigg( 1 - x_{\sq t}(q_2^2)\frac{m_\sq}{m_t}\bigg) + 2\mathrm{Li}_2\bigg( 1 - x_{\sq t}(q_1^2)\frac{m_t}{m_\sq}\bigg) - 2\mathrm{Li}_2\bigg( 1 - x_{\sq t}(q_1^2)\frac{m_\sq}{m_t}\bigg)\,.
\end{align*}
Then the hard matching coefficients are given by:
\begin{align*}
{\cal C}_{gg\to\gl\gl,3}^{(1)}&={\cal C}_{gg\to\gl\gl,4}^{(1)}={\cal C}_{gg\to\gl\gl,5}^{(1)}=0\\
{\cal C}_{gg\to\gl\gl,I}^{(1)}&=\mathrm{Re}\Bigg\{\frac{2N_c\pi^2}{3}+\frac{m_t^2+m_-^2n_f}{m_\gl^2} + 2\bigg[\gamma_E^2+\gamma_E \log\bigg(\frac{\mu_F^2}{4m_\gl^2}\bigg) - 2\bigg]N_c+ \gamma_g\log\bigg(\frac{\mu_R^2}{\mu_F^2}\bigg)\\
&+ \frac{m_+^2m_-^2}{2m_\gl^4}\log\bigg(\frac{m_-^2}{m_+^2}\bigg)\,n_l -\frac{(m_+^2 - m_t^2) (m_-^2 + m_t^2)^2}{2 m_\gl^4 (m_\gl^2 - (m_\sq + m_t)^2)}\,\beta_{\sq t}(m_\gl^2)  \log\big(x_{\sq t}(m_\gl^2)\big)\\
&+\frac{m_\gl^4 + 2 m_\gl^2 m_t ( m_t-m_\sq) - (m_\sq - m_t)^2 (m_\sq^2 - m_t^2)}{2 m_\gl^4}\,\beta_{\sq t}( - m_\gl^2) \log\big(x_{\sq t}( - m_\gl^2)\big)\\
&+\frac{m_-^2+m_t^2}{4(m_\sq^2-m_t^2)}\bigg[\frac{m_t^2}{m_\gl^2}F_3(-m_\gl^2,m_\gl^2,m_t,m_\sq)+\frac{m_\sq^2}{m_\gl^2}F_3(m_\gl^2,-m_\gl^2,m_\sq,m_t)\bigg]\\
&+\frac{n_l}{2}\bigg[\mathrm{Li}_2\bigg(\frac{m_\gl^2}{m_\sq^2}\bigg) - \mathrm{Li}_2\bigg(\! -\!\frac{m_\gl^2}{m_\sq^2}\bigg)\bigg]\frac{m_-^2}{m_\gl^2} +\bigg\{2+\gamma_E-\frac{\pi^2}{8}\bigg\}C_2(R_I)\\
&-\bigg\{\frac{m_\sq^2}{m_\sq^2-m_t^2}F_3(m_\gl^2,-m_\gl^2,m_\sq,m_t)+\frac{m_t^2(m_-^2+m_t^2)}{(m_\sq^2-m_t^2)(m_+^2-m_t^2)}F_3(-m_\gl^2,m_\gl^2,m_t,m_\sq)\\
&+ \frac{m_t^2}{m_+^2-m_t^2}\log^2\big(x_{tt}(4m_\gl^2)\big)+2n_l\bigg[\mathrm{Li}_2\bigg(\frac{m_\gl^2}{m_\sq^2}\bigg)-\mathrm{Li}_2\bigg(-\frac{m_\gl^2}{m_\sq^2}\bigg)\bigg]\,\bigg\}C'(R_I)\Bigg\}\ .
\end{align*}
Note that the representation $c_{gg\to\gl\gl,7}$ does not contribute in SU($3$). 

The hard matching coefficients for the $qg\to\sq\gl$ process also have a representation-dependent part that is not described by a simple quadratic Casimir invariant. In fact, the first representation, in Eq.~\eqref{eq:qgtoqgc1}, has an additional part compared to the representations in Eqs.~\eqref{eq:qgtoqgc2} and \eqref{eq:qgtoqgc3}, presumably due to the $s$-channel contribution that plays a role in the former. The hard matching coefficients are given by:
\begin{align*}
{\cal C}_{qg\to\sq\gl,I}&=\mathrm{Re}\Bigg\{\bigg[\frac{(m_\gl - m_\sq)N_c}{2m_{av}} + \frac{m_\gl^2-6m_\gl m_\sq+7m_\sq^2}{48m_{av}^2}C_F - \frac{m_\gl m_\sq}{8m_{av}^2N_c}\bigg]\,\pi^2+\gamma_g\log\bigg(\frac{\mu_R^2}{4m_{av}^2}\bigg)\\
& -\frac{\gamma_g+\gamma_q}{2}\log\bigg(\frac{\mu_F^2}{4m_{av}^2}\bigg)+\big(C_F+C_A\big)\bigg[\gamma_E^2+\gamma_E\log\bigg(\frac{\mu_F^2}{4m_{av}^2}\bigg)\bigg]+\frac{3(m_t^2 - m_\sq^2n_f)}{4m_\gl^2}\\
& +\bigg( \frac{4m_{av}^2}{m_-^2} + \frac{17(m_\sq^2 - 7m_\gl^2)}{24m_-^2}-\frac{m_+^2}{2m_\sq^2} \bigg)\,C_F - \frac{1}{6N_c}+\bigg[\frac{m_\gl^4 + 2m_\gl^2m_\sq^2 - 3m_\sq^4}{4m_\gl^4}n_l + \frac{5}{4N_c} \\
&+ \bigg(\frac{m_\sq}{m_\gl}- \frac{m_\gl^2(m_\gl^2 - 2m_\sq^2)}{2m_\sq^4}\bigg)C_F - \frac{(2m_\gl + m_\sq)N_c}{4m_\sq}\bigg]\log\!\bigg(\!1\!-\!\frac{m_\gl}{m_\sq}\!\bigg)+\bigg[\frac{(m_\gl^2 - 3m_\sq^2)^2n_l}{24m_\gl^4}\\
& +\frac{m_\gl^2(m_\gl^2 - 2m_\sq^2)C_F}{4m_\sq^4} + \frac{6m_\sq - 5m_\gl}{12m_\gl }N_c +\frac{(m_\gl - m_\sq)^2}{4m_\gl m_\sq N_c}\bigg]\log\bigg(\frac{m_\sq^2}{4m_{av}^2}\bigg)+\bigg[ \frac{m_\gl^3}{4m_-^2m_\sq N_c}\\
&+\bigg(\frac{m_\gl^2(m_\gl^2 - 2m_\sq^2)}{2m_\sq^4} + \frac{m_\gl}{2m_\sq} - \frac{m_\gl^3(m_\gl^2 + 3m_\sq^2)}{2m_\sq m_-^4}\bigg)\,C_F+ \frac{(4m_\gl^4-3m_\sq^4)N_c}{8m_-^4} + \frac{m_\gl N_c}{4m_\sq}\\
& + \frac{m_\gl^2m_\sq^2N_c}{m_-^4} +\frac{ 2m_\sq^4-11m_\gl^4}{8m_-^4N_c}- \frac{N_c}{24} \bigg]\log\bigg(\frac{m_\gl^2}{m_\sq^2}\bigg)+\bigg[ \frac{3 m_\gl^4 + 4 m_\gl^3 m_\sq - 4 m_\gl m_\sq^3 + 5 m_\sq^4}{16m_-^4N_c}\\
&+\bigg(\frac{3m_\gl^2 + m_\sq^2}{4m_-^2} + \frac{m_\sq(m_\gl^2 - 4m_\sq^2)}{8m_-^2m_{av}}\bigg)C_F \bigg]\log^2\bigg(\frac{m_\gl^2}{m_\sq^2}\bigg)+\bigg[\frac{2m_\sq^2m_-^2 - 2m_t m_\sq m_+^2}{m_\gl^2(m_\gl^2 - (m_\sq + m_t)^2)}\\
& + \frac{6m_tm_\sq (m_-^2 + m_t^2)+3 ( m_\sq^4 - m_t^4)}{4 m_\gl^4}  -\frac{m_\gl^2+10 m_\sq^2}{4 m_\gl^2}\bigg]\log(x_{\sq t}(m_\gl^2))\beta_{\sq t}(m_\gl^2)\\
&+\bigg[\frac{6m_\sq^2 - m_\gl^2}{24m_\gl^2} - \frac{3(m_\sq^2 - m_t^2)^2}{8m_\gl^4}\bigg]\log\!\bigg(\!\frac{m_t^2}{m_\sq^2}\!\bigg)+\bigg[\frac{2N_cm_+^2}{4m_{av}^2} + \frac{m_\gl m_\sq (3 N_c^2\!+\!1)}{4m_{av}^2N_c}\bigg]\mathrm{Li}_2\bigg(\!\frac{m_\sq}{m_\gl}\!\bigg)\\
&+\frac{(m_\gl-m_\sq)(N_c^2+1)}{4N_cm_{av}}\bigg[\mathrm{Li}_2\bigg(-\frac{m_\gl}{m_\sq}\bigg)-\mathrm{Li}_2\bigg(\frac{m_\sq}{m_\gl}\bigg)\bigg]-\bigg[\frac{m_\gl^2 + 2m_\gl m_\sq + 2m_\sq^2}{4m_{av}^2}C_F\\
& +\frac{m_-^2}{8m_{av}^2N_c}\bigg]\log\bigg(\frac{m_\gl^2}{m_\sq^2}\bigg)\log\bigg(1-\frac{m_\gl}{m_\sq}\bigg)+\bigg[\frac{m_\gl - m_\sq}{2m_{av}} - \frac{m_\sq^3}{m_-^2(m_\gl - m_\sq)}\bigg]\frac{F_6(m_\sq, m_\gl)}{2N_c}\\
& +\frac{(m_\gl - 2 m_\sq) (m_\gl^2 (1+3N_c^2)+ 2 m_\sq m_\gl (N_c^2+1))}{4m_-^4 N_c}\Big[m_\gl F_6(m_\gl,m_\sq)-m_\sq F_6(m_\sq,m_\gl)\Big]\\
&+\Bigg\{\bigg[ \frac{3m_\sq^2N_c}{8m_{av}^2}-\frac{m_\sq^2n_l}{6m_\gl m_{av}}-\frac{m_\gl(m_\gl^2 + 3m_\sq^2)}{48N_cm_\sq m_{av}^2}-\frac{(m_\gl^2 + 4 m_\sq^2)C_F}{24 m_{av}^2}\bigg]\,\pi^2+\frac{(m_\gl-m_\sq)C_F}{2m_{av}}\\
&+\frac{(m_\sq^2-m_t^2)^2}{2m_\gl^3m_\sq}\log\bigg(\frac{m_t^2}{m_\sq^2}\bigg)+\bigg[\frac{(m_\gl - m_\sq)C_F}{m_\sq} + \frac{N_c}{4} - \frac{m_\sq^2N_c}{8m_{av}^2}  +\frac{m_\gl(3m_\gl^2 + m_\sq^2)}{8N_cm_-^2m_{av}}\\
&- \frac{m_+^2}{4N_cm_\sq m_{av}}\bigg]\log\bigg(\frac{m_\gl^2}{m_\sq^2}\bigg)+\bigg[\frac{(N_c^2m_\gl - 2m_\sq)(3m_\gl + m_\sq)}{4N_cm_\gl m_{av}}-\frac{m_\sq^2(m_\gl - m_\sq)}{m_\gl^3}n_l\\
& -\frac{m_+^2(N_c^2 - 2)}{2N_cm_\sq m_\gl} - \frac{3m_\gl^2 - m_\sq^2}{4N_cm_\gl m_{av}} \bigg]\log\bigg(1-\frac{m_\gl}{m_\sq}\bigg)+\bigg[ \frac{(m_\gl^2 - 2m_\sq^2)(N_c^2 - 2)}{4N_cm_\sq m_\gl} -\frac{5m_\sq}{4N_cm_\gl}\\
& + C_F+\frac{1}{N_c}-\frac{m_\sq^3n_l}{2m_\gl^3}\bigg]\log\!\bigg(\frac{m_\sq^2}{4m_{av}^2}\bigg)+\bigg[\frac{m_\sq^2n_l}{2m_\gl m_{av}} + \frac{m_\gl^2 - m_\gl m_\sq - 4m_\sq^2}{4m_{av}^2}N_c \bigg]\,\mathrm{Li}_2\bigg(\!\frac{m_\sq}{m_\gl}\!\bigg)\\
&+\bigg[-\frac{m_\sq^2n_l}{4m_\gl m_{av}} + \frac{(3m_\sq - m_\gl)C_F}{2m_{av}} + \frac{(m_\gl - 2m_\sq)^2}{4N_cm_\sq m_{av}} +\frac{1}{2N_c}\bigg]\,\mathrm{Li}_2\bigg(-\frac{m_\gl}{m_\sq}\bigg)\\
&- \bigg[\frac{(m_\gl^2 - 3m_\sq^2)C_F}{4m_{av}^2} +\frac{m_\gl(m_\gl^2 - m_\sq^2N_c^2)}{8N_cm_\sq m_{av}^2}\bigg]\log\bigg(\frac{m_\gl^2}{m_\sq^2}\bigg)\log\bigg(1-\frac{m_\gl}{m_\sq}\bigg)\\
&+\bigg[\frac{m_\sq^2n_l}{16m_\gl m_{av}} + \frac{3(m_\gl^2 - 3m_\sq^2)(N_c^2 - 2)}{64N_cm_{av}^2}  - \frac{2m_\sq^2m_+^2}{N_cm_-^4} + \frac{m_\sq^2(m_\gl + 3m_\sq)^2}{8N_cm_-^4}\\
&- \frac{m_\gl m_\sq(3m_\gl + m_\sq)C_F}{16m_-^2m_{av}}+ \frac{(m_\gl - m_\sq)}{4N_cm_\sq} -\frac{4m_{av}^2}{8N_cm_-^2} +\frac{3m_\gl m_\sq m_+^2}{4N_cm_-^4}\bigg]\log^2\bigg(\frac{m_\gl^2}{m_\sq^2}\bigg)\\
&-\frac{m_\sq^2}{8 m_\gl m_{av}}F_3(m_\gl^2, - m_\gl m_\sq,m_\sq,m_t)+\frac{m_t^2}{8m_\gl m_{av}}F_3(m_\gl^2, - m_\gl m_\sq,m_t,m_\sq)\\
&-\frac{(m_\gl - 2m_\sq)m_\gl^2(m_\gl^2 - m_\gl m_\sq  +(N_c^2+2)m_\sq^2)}{2N_cm_\sq m_-^4}F_6(m_\gl,m_\sq)\\
&-\frac{(m_\gl - 2m_\sq)m_\sq(m_\gl^2 - m_\gl m_\sq(N_c^2+2)+m_\sq^2)}{2N_cm_-^4}F_6(m_\sq,m_\gl)\\
&-\frac{( (m_\sq - m_t)^2+m_\gl m_\sq) (m_\sq^2 - m_t^2) }{2m_\gl^2 m_\sq m_{av}}\log\big(x_{\sq t}(-m_\gl m_\sq)\big)\,\beta_{\sq t}(-m_\gl m_\sq)\\
&-\frac{( (m_\sq - m_t)^2-m_\gl^2) (m_\sq^2 - m_t^2)}{2m_\gl^3 m_{av}}\log\big(x_{\sq t}(m_\gl^2)\big)\,\beta_{\sq t}(m_\gl^2)\Bigg\}\bigg[1+\frac{2m_\gl C_FN_c}{m_\gl+m_\sq N_c^2}\delta_{I,1}\bigg]\\
&+\Bigg\{\frac{m_\gl^2(m_\gl - 2m_\sq)}{2m_-^4}\Big[m_\sq F_6(m_\sq, m_\gl) - m_\gl F_6(m_\gl, m_\sq)\Big] -\frac{m_\sq m_+^2}{16m_-^2m_{av}}\log^2\bigg(\frac{m_\gl^2}{m_\sq^2}\bigg)+2\\
& - \frac{m_\gl - m_\sq}{2m_{av}}\bigg[\mathrm{Li}_2\bigg(\frac{ m_\sq}{m_\gl}\bigg) +\frac{1}{2}\log\bigg(\frac{m_\gl^2}{m_\sq^2}\bigg) + \frac{\pi^2}{12}\bigg]+ \gamma_E  -\frac{m_\gl(m_\gl - m_\sq)}{8m_{av}^2}\pi^2 \Bigg\}C_2(R_I)\Bigg\}\,,
\end{align*}
where $\delta_{I,1}=1$ for the first representation and vanishes for the other representations. Also, we have defined the function:
\begin{align*}
F_6(m_\sq, m_\gl) &= \mathrm{Li}_2\bigg( 2 - \frac{m_\sq}{m_\gl} \bigg) - \mathrm{Li}_2\bigg(1- \frac{m_\gl}{m_\sq} + \frac{m_\gl ^2}{m_\sq^2}\bigg)\,.
\end{align*}

\end{fmffile}

\bibliographystyle{JHEP}
\providecommand{\href}[2]{#2}\begingroup\raggedright\endgroup

\end{document}